\documentclass{article}

\usepackage{microtype}
\usepackage{graphicx}
\usepackage{subcaption}
\usepackage{booktabs} 

\usepackage{hyperref}

\usepackage[accepted]{icml2026}

\usepackage{amsmath}
\usepackage{amssymb}
\usepackage{mathtools}
\usepackage{amsthm}

\usepackage[capitalize,noabbrev]{cleveref}

\theoremstyle{plain}

\theoremstyle{definition}

\theoremstyle{remark}

\usepackage[textsize=tiny]{todonotes}

\usepackage{xspace}
\newcommand{\consensustester}{Agora\xspace} 
\newcommand{\agenta}{Orchestrator\xspace} 
\newcommand{\agentb}{Strategy\xspace} 
\newcommand{\agentc}{TestGen\xspace} 
\newcommand{\bfagenta}{\textbf{Orchestrator}\xspace} 
\newcommand{\bfagentb}{\textbf{Strategy}\xspace} 
\newcommand{\bfagentc}{\textbf{TestGen}\xspace} 
\usepackage{tikz}
\newcommand{\circled}[1]{\tikz[baseline=(char.base)]{
  \node[shape=circle,draw,inner sep=0.2pt,minimum size=0.8em, font=\scriptsize] (char) {#1};
}}
\newcommand{\BugAnalysisBox}[3]{%
  \begin{tcolorbox}[
    colframe=orange!60!black,    
    coltitle=orange!80!black,     
    colbacktitle=orange!20!white, 
    colback=white, 
    title={#1}, 
    fonttitle=\bfseries,
    nobeforeafter, center title,
    enhanced,breakable
  ]
    \textbf{Analysis}: #2 
  \end{tcolorbox}
}
\usepackage[most]{tcolorbox}

\newcommand{\PatternBox}[3]{%
  \begin{tcolorbox}[
    enhanced,
    breakable,
    colframe=blue!50!black,    
    colback=white,             
    coltitle=white,            
    colbacktitle=blue!50!black, 
    title={#1}, 
    fonttitle=\bfseries\sffamily, 
    sharp corners,             
    boxrule=0.8pt,             
    titlerule=0pt,             
    center title,              
    left=10pt, right=10pt, top=10pt, bottom=10pt 
  ]
    #2 
  \end{tcolorbox}
}

\usepackage[table]{xcolor}

\usepackage[most]{tcolorbox} 
\usepackage{xcolor}          
\usepackage{fancyvrb} 
\definecolor{SeaCornSoft}{HTML}{FFF8DC} 
\definecolor{RoyalAquaDeep}{HTML}{006064}
\newtcolorbox{agentbox}[1][]{
  enhanced, breakable, drop shadow,
  width=\linewidth, boxrule=0.8pt,
  colback=SeaCornSoft,       
  colframe=RoyalAquaDeep,    
  left=4pt,right=4pt,top=4pt,bottom=4pt,
  fonttitle=\small,
  title={\bfseries #1}
}
\definecolor{DeepRed}{RGB}{139,30,30}
\definecolor{LightRed}{RGB}{242,198,198}
\definecolor{DeepBlue}{RGB}{31,78,121}
\definecolor{LightBlue}{RGB}{214,228,240}

\icmltitlerunning{\consensustester: Toward Autonomous Bug Detection in Production-Level Consensus Protocols with LLM Agents}

\begin{document}

\twocolumn[
\icmltitle{
    \begin{tabular}{cc}
        \raisebox{-15pt}{\includegraphics[height=35pt]{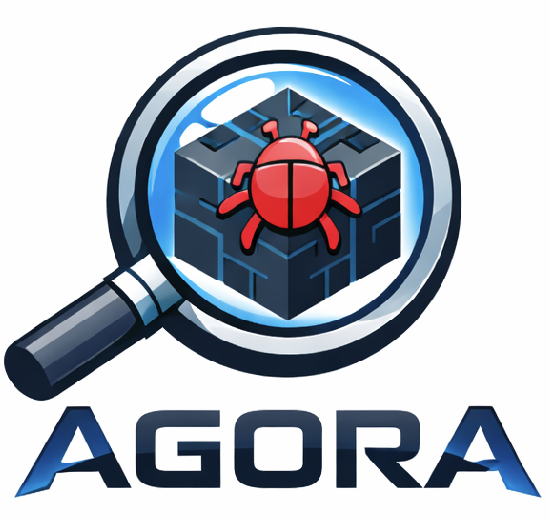}} & 
        \begin{tabular}{c}
             \consensustester: Toward Autonomous Bug Detection in 
            Production-Level\\ Consensus Protocols with LLM Agents
        \end{tabular}
    \end{tabular}
}






  \icmlsetsymbol{equal}{*}

\begin{icmlauthorlist}
\icmlauthor{Xiang Liu}{equal,nus}
\icmlauthor{Sa Song}{equal,bupt}
\icmlauthor{Zhaowei Zhang}{pku}
\icmlauthor{Huiying Lan}{nus}
\icmlauthor{Jason Zeng}{0g}
\icmlauthor{Ming Wu}{0g}
\icmlauthor{Michael Heinrich}{0g}
\icmlauthor{Yong Sun}{bupt}
\icmlauthor{Ceyao Zhang}{pku}
\end{icmlauthorlist}

\icmlaffiliation{nus}{School of Computing, National University of Singapore}
\icmlaffiliation{bupt}{School of Information and Telecommunication Engineering, Beijing University of Posts and Telecommunications}
\icmlaffiliation{pku}{Peking University}
\icmlaffiliation{0g}{0G Labs}

\icmlcorrespondingauthor{Yong Sun}{sunyong@bupt.edu.cn}
\icmlcorrespondingauthor{Ceyao Zhang}{ceyaozhang@pku.edu.cn}
  \icmlkeywords{Multi-agent}

  \vskip 0.3in

]



\printAffiliationsAndNotice{\icmlEqualContribution}

\begin{abstract}
Consensus protocols form the backbone of distributed systems and blockchains, where implementation bugs can cause data corruption and financial losses. 
While LLM-based approaches show promise in code analysis, they struggle 
with deep protocol-level logic bugs involving complex state-dependent behaviors across multiple execution stages.
We present \consensustester, a domain-aware multi-agent framework that integrates hypothesis-driven testing with LLM capabilities for systematic protocol verification. \consensustester employs specialized agents that collaboratively explore protocol state spaces, synthesize attack scenarios using domain-specific constraints, and validate findings through iterative refinement. This explicit role separation enables reasoning about global protocol invariants beyond single-function code analysis.
We evaluate \consensustester on four consensus implementations (Raft, EPaxos, HotStuff, BullShark) using four state-of-the-art LLMs. \consensustester discovers 15 previously unknown protocol-level logic bugs that violate safety properties, while existing LLM-based agents fail to detect any such protocol-level logic bugs. Our results demonstrate that domain-aware multi-agent collaboration is essential for detecting deep logic bugs in complex protocols.
\end{abstract}

\section{Introduction}
\label{sec:intro}

Consensus protocols, including Crash Fault-Tolerant (CFT)~\cite{Lamport1998ThePP,Lamport2001PaxosMS,Moraru2013ThereIM,Ongaro2014InSO} and Byzantine Fault-Tolerant (BFT)~\cite{Castro1999PracticalBF, Lamport2011ByzantizingPB, Yin2018HotStuffBC,Danezis2021NarwhalAT} protocols, serve as the cornerstone of modern distributed systems, ensuring consistency and availability in the presence of faults. Among these, the CFT protocols are widely adopted in distributed databases~\cite{taft2020cockroachdb}, storage systems~\cite{corbett2013spanner}, and cluster coordination services~\cite{hunt2010zookeeper} to guarantee the correctness of online transactions and data storage. Conversely, the BFT protocols form the backbone of blockchain systems~\cite{nakamoto2008bitcoin}, designed to secure transactions against malicious actors. 

Consequently, any defects in the implementation of these consensus protocols can lead to severe consequences, ranging from data corruption to significant financial losses~\cite{zheng2017paxosstore}.
Despite the critical importance of these systems, bug detection for distributed protocol implementations remains a formidable challenge and has long been a focal point of research. The inherent complexity of distributed systems, coupled with the intricate logic of consensus algorithms, makes subtle bugs notoriously difficult to detect~\cite{Ongaro2014InSO}. Even widely used and mature libraries (e.g., etcd~\cite{etcdioraft}), which have been under development for many years, are still found to contain critical bugs that require experienced engineers extensive periods to uncover and resolve~\cite{Zhou2021FoundationDBAD}.

\begin{figure*}[!th]
  \centering
  \includegraphics[width=\textwidth]{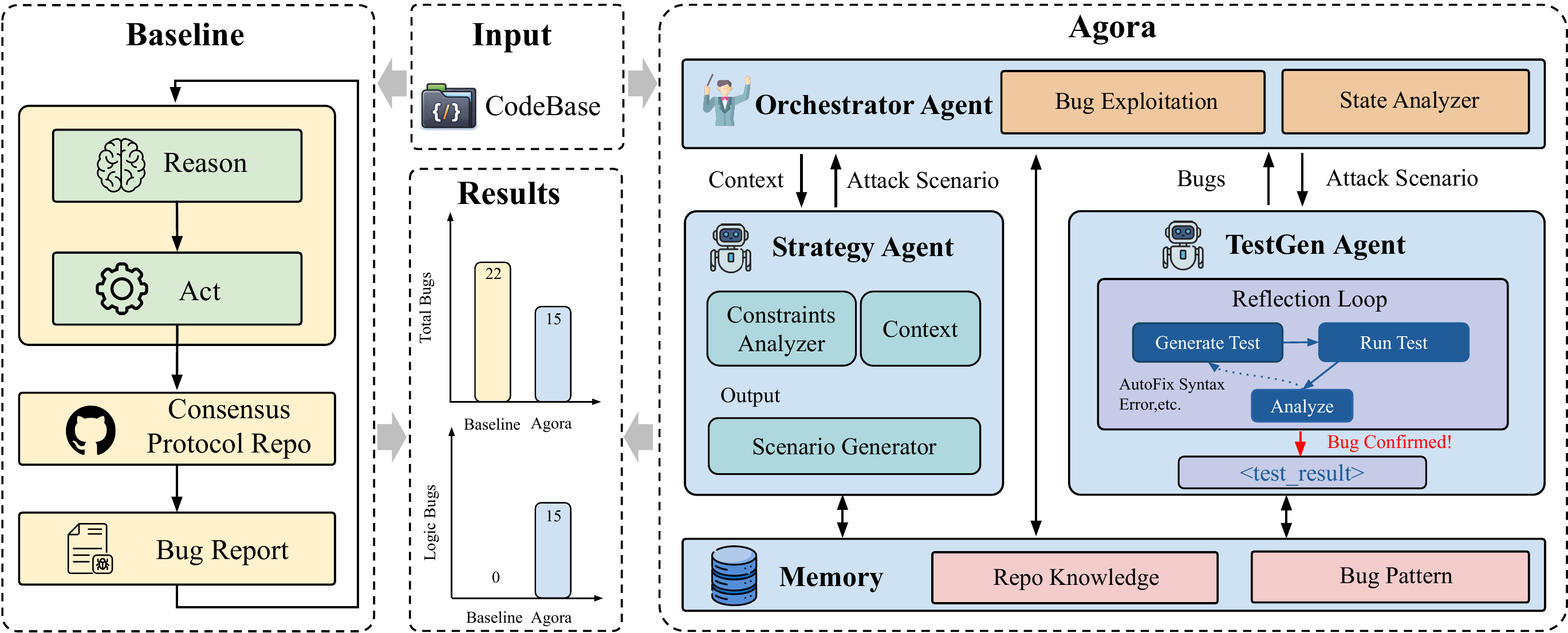}
  \caption{\textbf{Left} is the baseline overview. \textbf{Middle} shows the input and result comparison where baselines could only detect \textbf{low-level implementation bugs}, but \consensustester could detect critical \textbf{protocol-level logic bugs}. \textbf{Right} is the overview of \consensustester, which consists of three agents: \agenta agent, \agentb agent and \agentc agent.}
  \label{fig:overview}
\end{figure*}

In recent years, the rapid advancement of Large Language Models (LLMs)~\cite{Achiam2023GPT4TR,DeepSeekAI2025DeepSeekR1IR} and autonomous agent systems~\cite{yao2022react,cheng2024exploring} has revolutionized software engineering tasks, helping engineers save significant time and rapidly perform tasks such as code analysis~\cite{feng2020codebert} and bug detection~\cite{chen2021evaluating}. While LLMs and agents have demonstrated strong capabilities in identifying common coding errors and vulnerabilities~\cite{Kaniewski2025ASL}, their effectiveness remains limited: they primarily excel at detecting superficial or localized bugs, but struggle to reason about the complex, state dependent logic inherent in distributed consensus protocols~\cite{Wang2024SanitizingLL,Wang2024LLMDFAAD}.
Therefore, to fully leverage the general code understanding capabilities of LLMs and efficiently detect bugs in consensus protocols and their implementations, it is necessary to integrate domain knowledge of distributed consensus with multi-agent systems~\cite{Yang2025KernelGPTEK,Yang2025KNighterTS}. Such a fusion enables the identification of subtle errors that may otherwise lead to severe data inconsistencies or substantial economic losses, thereby ensuring system correctness and security for researchers and engineers.

Therefore, in this paper, we first analyze bugs in consensus protocols and identify corresponding patterns, and summarize the constraints under which bugs occur in consensus protocols. Then, based on the testing principle of hypothesis-driven testing (HDT)~\cite{Kell2004HereIT}, we are the first to construct a multi-agent system \consensustester to better integrate domain knowledge with the strong code understanding and generalization capabilities of LLMs to analyze consensus protocol code repositories and efficiently discover bugs in consensus protocols. Specifically, as shown in Figure~\ref{fig:overview}, our \consensustester is composed of three agents to realize the decoupling of workflow control, consensus scenarios properties, and test generation, enabling efficient deep bug hunting in consensus protocols:
\textit{(1)} \bfagenta agent is responsible for agent interactions and summarizing existing bugs, \textit{(2)} \bfagentb agent is responsible for generating special consensus protocol bug scenarios, \textit{(3)} \bfagentc agent generates corresponding unit tests based on the bug scenarios to expose bugs.
We also used our \consensustester system to conduct extensive experiments on the repositories of four popular and widely used research- and production-level consensus protocols to validate the effectiveness of our approach. Our main contributions are summarized as follows:

\begin{figure*}[!th]
  \centering
  \includegraphics[width=\textwidth]{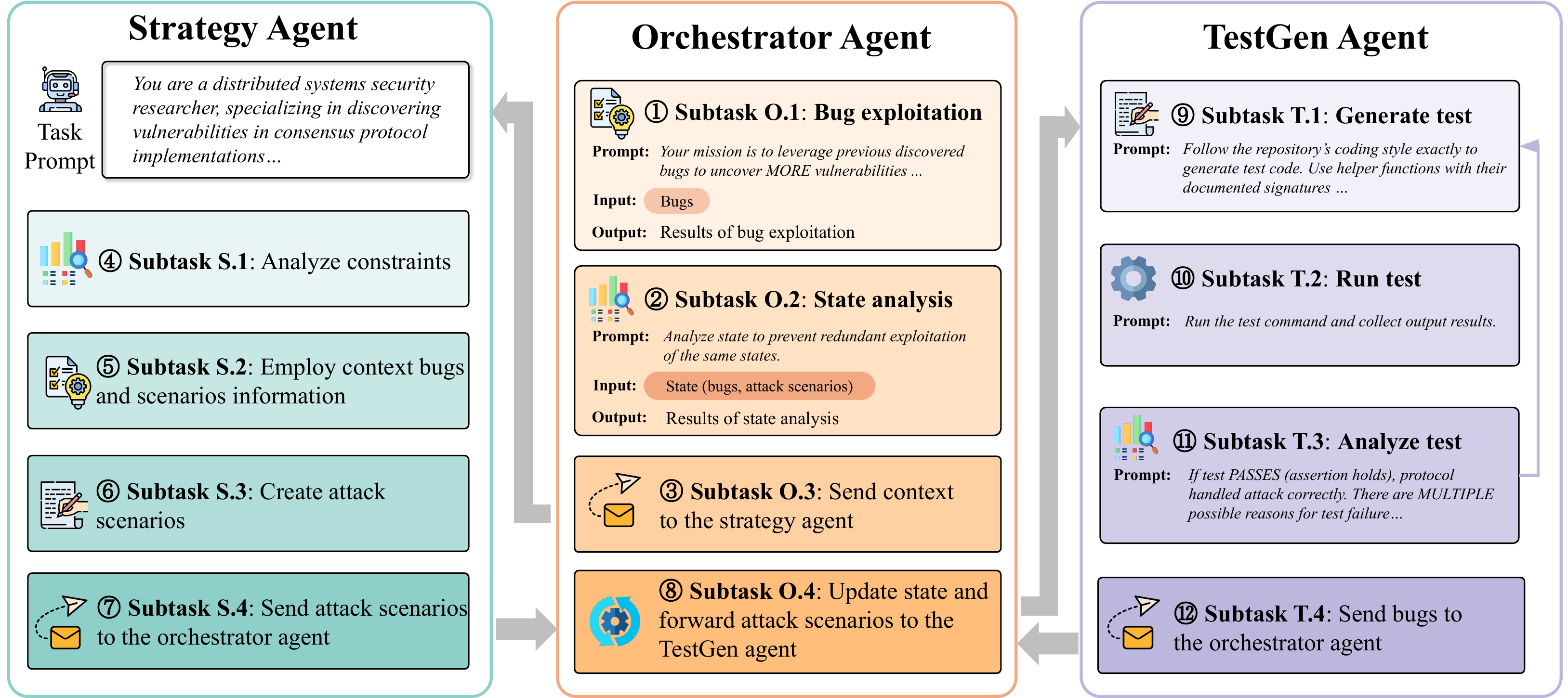}
  \caption{The overview of \consensustester workflow, which consists of three agents and twelve steps/subtasks (following the time order, from Step \circled{1} to Step \circled{12}). \bfagenta agent is responsible for bug exploitation, coordinating the agents, maintaining global state, and sending and receiving messages among agents; \bfagentb agent is responsible for generating attack scenarios based on the analysis results and the information it receives; \bfagentc agent is responsible for generating valid unit tests through a reflection loop to detect bugs in consensus systems. Each agent consists of four subtasks.}
  \label{fig:overview2}
\end{figure*}

\begin{itemize}
\item We are the first to combine domain knowledge of distributed consensus with LLM-based agents to detect the protocol-level logic bugs (\S\ref{sec:goal}).

\item We implement a flexible agent framework, \consensustester, which effectively follows the popular hypothesis-driven testing (HDT) paradigm and employs three agents (\agenta, \agentb, and \agentc) to automatically perform code bug detections for multiple consensus protocol code repositories (\S\ref{sec:architecture}).

\item We evaluate \consensustester using multiple LLM models and conduct extensive experiments on four popular consensus protocol repositories implemented in different programming languages. Our approach discovers 15 zero-day bugs, demonstrating that \consensustester can significantly improve the safety of distributed systems  (\S\ref{sec:expoverall}). 

\item We conduct comparative analysis and systematic ablation studies, demonstrating that domain-aware multi-agent collaboration is essential: removing a single component reduces effectiveness by 73--100\%, and each design choice contributes critically to performance (\S\ref{sec:ablation}).
\end{itemize}

\section{Preliminary and Related Works}
\label{sec:pre}

\subsection{Hypothesis-Driven Testing}

Compared with traditional testing, which only focuses on verifying software does what it is supposed to do, hypothesis-driven testing (HDT)~\cite{Kell2004HereIT} aims to explore under what conditions the software may fail. By explicitly formulating vulnerability hypotheses, HDT enables higher code coverage and more systematic bug discovery. A vulnerability hypothesis is defined as a tuple $H = (C, A, E, O)$, where $C$ denotes the preconditions required to trigger a vulnerability, $A$ represents the action sequence that may activate it, $E$ specifies the expected buggy behavior, and $O$ defines the oracle assertions used to validate the bug.
Therefore, the goal of \consensustester is to construct an automated loop that follows the HDT paradigm and continuously discovers bugs by generating, instantiating, and validating vulnerability hypotheses.

\subsection{Consensus Protocols}

Consensus protocols mainly fall into two categories: Crash Fault-Tolerant (CFT) protocols, such as Paxos~\cite{Lamport1998ThePP,Lamport2001PaxosMS}, Fast Paxos~\cite{Lamport2006FastP}, EPaxos~\cite{Moraru2013ThereIM}, and Raft~\cite{Ongaro2014InSO}; and Byzantine Fault-Tolerant (BFT) protocols, such as PBFT~\cite{Castro1999PracticalBF}, HotStuff~\cite{Yin2018HotStuffBC}, DAG-Rider~\cite{Keidar2021AllYN}, Narwhal–Tusk~\cite{Danezis2021NarwhalAT}, and BullShark~\cite{Giridharan2022BullsharkDB}.
CFT consensus protocols are primarily used in traditional banking systems and Internet companies for data transactions and cluster coordination, whereas BFT consensus protocols form the foundation of blockchain systems and on-chain transactions. These protocols are widely used in both research and production settings, with code repositories implemented in multiple programming languages, such as Go and Rust. The distinction is particularly important because BFT systems explicitly tolerate malicious behavior from participating nodes, whereas CFT systems do not; as a result, analyzing these two classes of protocols involves fundamentally different \textbf{Constraints}.

To analyze bugs in consensus protocols, it is essential to summarize \textbf{bug patterns} previously identified through extensive human expertise, in order to more effectively uncover bugs that violate system safety and correctness and may lead to erroneous transactions and data corruption (i.e., protocol-level \textbf{Logic Bugs}), rather than low-level \textbf{Implementation Bugs} such as out-of-memory errors, memory leaks, integer overflows, or simple logical flaws.

The distributed protocols are primarily required to satisfy two fundamental properties: liveness and safety. Only the causes that lead to violations of these two properties constitute true protocol-level logic bugs. Such causes can be broadly categorized into five classes: (i) Recovery\&Execution Divergence, (ii) Persistence\&Monotonicity Violations, (iii) Dependency\&Topology Flaws, (iv) Message Binding\&Signature Violations, and (v) Resource\&Operational Visibility Violations.

\subsection{LLM-based Multi-Agent System}
The paradigm of LLM applications focuses on multi-agent architectures, which decompose complex tasks into collaborative, specialized roles. In this way, multi-agent systems enable more focused reasoning, mitigate hallucinations, and manage long-horizon objectives more effectively than single-agent models~\cite{Chen2023AgentVerseFM,Guo2024LargeLM,Yang2025AgentNetDE}. The efficacy of this paradigm also enables structured coordination mechanisms and communication protocols (see the formalization in Appendix~\ref{appendsec:multiagent}). 

Early dual-agent role-playing systems, such as CAMEL~\cite{Li2023CAMELCA} and ProAgent~\cite{Zhang2023ProAgentBP}, demonstrate that collaboration among agents can elicit step-by-step reasoning. More recent frameworks extend this idea by assigning explicit roles to multiple agents: MetaGPT~\cite{hong2023metagpt} and ChatDev~\cite{Qian2023ChatDevCA} employ manager, designer, and coder agents to simulate software engineering workflows; Magnetic-One~\cite{Fourney2024MagenticOneAG} and AG2~\cite{AG2} introduce a central coordinator that assigns tasks to different workers. OWL~\cite{Hu2025OWLOW} further decouples multi-agent systems, enabling effective transfer across domains to solve new tasks.
However, existing multi-agent systems still lack the ability to reason deeply about code patterns, especially to integrate domain knowledge of consensus protocol systems for efficiently discovering deep bugs.

\begin{algorithm}[!t]
\caption{\consensustester Workflow}
\label{alg:workflow}
\begin{algorithmic}[1]
    \STATE {\bfseries Input:} Repository Knowledge $\mathcal{R}$, Bug Patterns $\mathcal{P}$, Protocol Constraints $Con$
    \STATE {\bfseries Output:} Detected Bugs $\mathcal{B}$
    \STATE $\mathcal{S} \leftarrow \emptyset$ \COMMENT{Initialize global state}
    
    \WHILE{exploration budget remains}
        \STATE \textit{// --- Orchestrator Agent ---}
        \STATE $\mathcal{BE} \leftarrow \text{bug-exploitation}(\mathcal{B})$ \COMMENT{Analyze historical bugs found}
        \STATE $Z \leftarrow \text{state-analyzer}(\mathcal{S})$ \COMMENT{Summarize global state}
        
        \STATE \textit{// --- Strategy Agent ---}
        \STATE $AttS \leftarrow \text{Strategy}(\mathcal{BE}, Z, Con, \mathcal{P}, \mathcal{R})$ \COMMENT{Synthesize attack scenarios}
        \STATE \textbf{SendToOrchestrator}($AttS$) \COMMENT{Update $\mathcal{S}$}
        
        \STATE \textit{// --- TestGen Agent ---}
        \REPEAT
            \STATE $T \leftarrow \text{GenerateUnitTests}(AttS, \mathcal{R})$ \COMMENT{Repo knowledge-based gen unit test}
            \STATE $R \leftarrow \text{ExecuteAndAnalyze}(T)$
        \UNTIL{$R = \text{Success} \lor \text{MaxRetries}$}
        
        \IF{$R = \text{Success}$}
            \STATE \textbf{SendToOrchestrator}($T$) \COMMENT{Update $\mathcal{S}$}
        \ENDIF
    \ENDWHILE
\end{algorithmic}
\end{algorithm}

\subsection{LLM-powered Bug Detection} 

Recent advances in LLMs have significantly influenced automated bug detection and program analysis. Prior work has demonstrated that LLMs can assist developers in tasks such as bug localization, program repair, test generation, and vulnerability detection by leveraging their strong code understanding and natural language reasoning capabilities~\cite{feng2020codebert,chen2021evaluating,pearce2022asleep}.

LLMSAN~\cite{Wang2024SanitizingLL}, SCALE~\cite{Wen2024SCALECS}, LLMDFA~\cite{Wang2024LLMDFAAD} and PrimeVul~\cite{Ding2025VulnerabilityDW} all leverage LLMs to assist in bug detection. KNighter~\cite{Yang2025KNighterTS} is the first to employ an agent-based system for analyzing bugs in LLM system code.
However, these existing LLM-powered bug detection techniques largely focus on localized code patterns (\textbf{implementation bugs}) and lack the ability to reason about complex, state-dependent behaviors that span multiple components or execution stages (\textbf{logic bugs}). Moreover, these approaches do not leverage multi-agent systems to combine with HDT paradigm to achieve more comprehensive bug discovery.

\section{Method}
\label{sec:method}


\subsection{Motivation and Design Goal}
\label{sec:goal}

We introduce \consensustester, which integrates domain knowledge of distributed consensus with an LLM-based multi-agent system to address the fundamental challenge of comprehensively detecting protocol-level logic bugs through HDT. Beyond this core capability, \consensustester is designed with several desirable properties. It is \textbf{{effective}}, capable of uncovering deep logic bugs; it provides  \textbf{{interpretability}}, producing detailed analysis reports for discovered bugs to facilitate debugging; it is \textbf{{efficient}}, requiring less communication and fewer tokens and yielding a low false-positive rate; it is  \textbf{{complete}}, applicable to both CFT and BFT systems; and it is \textbf{{flexible}}, featuring a modular, decoupled design that supports plug-and-play extension to other domains. 

Moreover, our method should prioritize using computational resources to discover important \textbf{logic bugs}, rather than wasting tokens on \textbf{implementation bugs}.

\begin{table*}[!th]
    \centering
    \caption{Overall results for the baselines (top of the table) and \consensustester (bottom of the table). 
    In total, the former correspond to implementation bugs ($\downarrow$, fewer is better, with \textcolor{DeepBlue}{darker blue} indicating worse results), while the latter correspond to logic bugs ($\uparrow$, more is better, with \textcolor{DeepRed}{darker red} indicating better results). According to \S\ref{sec:goal}, fewer implementation bugs are preferred, whereas more logic bugs are desirable. The best results are highlighted in \textbf{bold}.}
    \label{tab:overall}
    \resizebox{\textwidth}{!}{
    \begin{tabular}{l|cccc|cccc|c}
        \toprule
        \textbf{Bug Type}
        & \multicolumn{4}{c}{Implementation Bug ($\mathbf{\downarrow}$) }
        & \multicolumn{4}{c}{\textbf{Logic Bug} ($\mathbf{\uparrow}$)}
        &  \\
        \cmidrule(lr){2-5} \cmidrule(lr){6-9} \cmidrule(lr){10-10}
         \textbf{Consensus Protocol} & \textbf{Raft} & \textbf{EPaxos} & \textbf{HotStuff} & \textbf{BullShark}
        & \textbf{Raft} & \textbf{EPaxos} & \textbf{HotStuff} & \textbf{BullShark}
        & \textbf{Total} \\
        \midrule
        \midrule
        GPT-5.2~\cite{openai2025gpt52} & \cellcolor{DeepBlue!40}4 & \cellcolor{DeepBlue!60}7 & \cellcolor{DeepBlue!80}11 & 0 & 0 & 0 & 0 & 0 & 22/0 \\
        Gemini 3.0 Pro Preview~\cite{google2025gemini3} & 0 & 0 & 0 & 0 & 0 & 0 & 0 & 0 & 0/0 \\
        Claude Sonnet 4.5~\cite{anthropic2025claude45} & 0 & 0 & 0 & 0 & 0 & 0 & 0 & 0 & 0/0 \\
        Qwen3 Coder 480B A35B~\cite{qwen2025qwen3} & 0 & \cellcolor{DeepBlue!20}3 & 0 & 0 & 0 & 0 & 0 & 0 & 3/0 \\
        \midrule
        \rowcolor{gray!15} \textbf{Total (Baselines)} & 4 & 7 & 11 & 0 & 0 & 0 & 0 & 0 & 22/0 \\
        \midrule
        \midrule
        GPT-5.2~\cite{openai2025gpt52} & 0 & 0 & 0 & 0 & \cellcolor{DeepRed!30}1 & \cellcolor{DeepRed!60}3 & \cellcolor{DeepRed!60}3 & \cellcolor{DeepRed!30}1 & 0/8 \\
        Gemini 3.0 Pro Preview~\cite{google2025gemini3} & 0 & 0 & 0 & 0 & \cellcolor{DeepRed!30}1 & \cellcolor{DeepRed!90}6 & \cellcolor{DeepRed!60}3 & \cellcolor{DeepRed!30}1 & \textbf{0/11} \\
        Claude Sonnet 4.5~\cite{anthropic2025claude45} & 0 & 0 & 0 & 0 & 0 &\cellcolor{DeepRed!70}4 & \cellcolor{DeepRed!30}1 & \cellcolor{DeepRed!30}1 & 0/6 \\
        Qwen3 Coder 480B A35B~\cite{qwen2025qwen3} & 0 & 0 & 0 & 0 & \cellcolor{DeepRed!30}1 &\cellcolor{DeepRed!90} 6 & \cellcolor{DeepRed!45}2 & 0 & 0/9 \\
        \midrule
         \rowcolor{blue!10} \textbf{Total (\consensustester)} & 0 & 0 & 0 & 0 & 1 & 9 & 4 & 1 & \textbf{0/15} \\
        \bottomrule
    \end{tabular}
    }
\end{table*}

\subsection{Architecture}
\label{sec:architecture}

As shown in Figure~\ref{fig:overview}, \consensustester is comprised of three core agents:
(i) \bfagenta agent is responsible for workflow control and consists of two components. The \texttt{bug-exploitation} component leverages bugs previously discovered by \consensustester to more effectively trigger new bugs, while the \texttt{state-analyzer} component maintains the states of the other two agents and prevents \consensustester from repeatedly exploring identical bug scenarios;
(ii) \bfagentb agent contains two components: a \texttt{constraints-analyzer} that analyzes the constraints of bug attack scenarios and supports both CFT and BFT protocols (\textbf{complete}), and an attack \texttt{scenario-generator} that produces protocol-specific scenarios to facilitate the discovery of deep logic bugs;
(iii) \bfagentc agent employs a \texttt{reflection-loop} component to automatically generate, execute, and analyze unit tests, reducing false positives and synthesizing user-friendly bug reports based on the generated attack scenarios  (\textbf{interpretability}).

Overall, this design, where one agent coordinates the workflow, one agent identifies protocol-level logic bug scenarios, and one agent focuses on test generation, effectively decouples system functionality, is composed of five components and enables \textbf{flexibility} and extensibility. \S\ref{sec:exp} demonstrates the \textbf{efficiency} and \textbf{effectiveness} of \consensustester.

\subsection{Knowledge Library}
\label{sec:knowledgelibrary}

For \texttt{bug patterns}, we crawl GitHub repositories of consensus protocols to collect issues that have been confirmed as bugs and summarize them accordingly, filtering out low-level implementation bugs. In this way, we incorporate domain-specific knowledge of consensus protocols to help \consensustester understand protocol-level logic bugs and generate more effective attack scenarios.
For \texttt{constraints}, we summarize the different settings and design choices of consensus protocols and, combined with the analysis of bug patterns, derive the distinct conditions under which bugs are triggered in CFT and BFT systems. For example, assuming malicious behavior of nodes in a CFT system can easily violate safety properties; however, such assumptions are impractical in real deployments and would unnecessarily waste computational resources.

By combining bug patterns and constraints, we introduce a succinct knowledge library for \consensustester that only provides the necessary domain knowledge for \consensustester. The details of bug patterns are presented in Appendix~\ref{appendsec:bugpattern} and the details of constraints are in Appendix~\ref{appendsec:constraints}.

\subsection{Memory}
\label{sec:memory}

In our system, the memory maintained by \consensustester consists only of repository knowledge and bug patterns. Bug patterns are described in detail in \S\ref{sec:knowledgelibrary}. Repository knowledge typically includes the repository name, the type of consensus protocol, protocol settings, implementation language, and testing methodologies. This information enables \consensustester to efficiently support consensus protocols implemented in different programming languages and with diverse designs.

Both repository knowledge and bug patterns are shared across the agents of \consensustester for analysis and are lightweight and easy to maintain (the details are in Appendix~\ref{appendsec:memory}). Therefore, we store them in the system memory and maintain them within the execution environment of \consensustester LLM-based multi-agent system.

\subsection{Workflow Details}

Before reaching the termination condition or exhausting the budget, as shown in Figure~\ref{fig:overview2}, the workflow of \consensustester consists of 12 steps (subtasks).

In Step \circled{1}, \agenta performs its subtask O.1 and conducts bug exploitation using prompts. By analyzing previously discovered bugs, it provides better guidance for \agentb to derive new deep bugs based on prior bugs found from \agentc.
In Step \circled{2}, \agenta performs subtask O.2 and uses prompts to analyze attack scenarios proposed earlier by \agentb together with bugs discovered by \agentc, producing a global state.
In Step \circled{3}, \agenta performs subtask O.3 and sends the context information, i.e., the results of bug exploitation and state analysis, to \agentb.

In Step \circled{4}, \agentb performs subtask S.1 and analyzes protocol constraints using prompts.
In Step \circled{5}, \agentb performs subtask S.2 and employs the context from bug exploitation and state analysis.
In Step \circled{6}, \agentb performs subtask S.3 and, using prompts together with the results from Steps \circled{4} and \circled{5}, generates attack scenarios. Specifically, for a given consensus protocol, it simulates scenarios by controlling node behaviors such as joining, going offline, crashing, or recovering, as well as manipulating message sending orders and conflict relationships to produce bug-prone scenarios.
In Step \circled{7}, \agentb completes its subtask S.4 and sends the generated attack scenarios to \agenta.
In Step \circled{8}, \agenta completes its subtask O.4, updates the global state, and forwards the attack scenarios to \agentc.

In Step \circled{9}, \agentc performs its subtask T.1 and generates unit tests based on the received attack scenarios using prompts.
In Step \circled{10}, \agentc performs subtask T.2 and executes the generated unit tests using prompts.
In Step \circled{11}, \agentc performs subtask T.3 and analyzes the generated unit tests together with the execution results from Step \circled{10} using prompts. If successful, this indicates that a logic bug in the consensus protocol has been discovered, and the workflow proceeds to Step \circled{12}. Otherwise, the process returns to Step \circled{9}. The number of iterations is bounded by a maximum value, \texttt{num\_iter}; reaching this limit indicates that the current attack scenario is invalid and should be regenerated.
In Step \circled{12}, \agentc completes its subtask T.4 and reports the successfully discovered bug to \agenta. 
It is worth noting that although Steps \circled{9}–\circled{11} are described as separate subtasks, they can be implemented using a single prompt that enables the \agentc agent to sequentially complete the three subtasks in the reflection loop.

We revisit HDT with \consensustester. In our framework, the attack scenarios correspond to the preconditions $C$, the unit tests generated by \agentc constitute the action sequence $A$, and the expected buggy behavior $E$ and the oracle assertions $O$ are contained in Steps \circled{10} and \circled{11} of \agentc. As a result, \consensustester effectively integrates domain knowledge of consensus protocols and the HDT paradigm into a multi-agent system.
In summary, \agenta, \agentb, and \agentc are each composed of four subtasks. The prompt details of our agents are provided in Appendix~\ref{appendsec:prompt}. The overall workflow is illustrated in Algorithm~\ref{alg:workflow}.

\begin{table}[!t]
    \centering
    \caption{Results for bug detection in ablation studies. We compare the original \consensustester with variants in which the corresponding components are removed.}
    \label{tab:ablation_study}
\resizebox{\linewidth}{!}{
    \begin{tabular}{lccccc}
        \toprule
        \textbf{Methods}
        & \textbf{Raft} & \textbf{EPaxos} & \textbf{HotStuff} & \textbf{BullShark}
        & \textbf{Total} \\
        \midrule
        \consensustester & 1 & 6 & 3 & 1 & 11 \\
        w/o bug-exploitation & 0 & 0 & 3 & 0 & 3 \\
        w/o state-analyzer & 0 & 0 & 0 & 0 & 0 \\
        w/o constraints-analyzer & 0 & 0 & 1 & 0 & 1 \\
        w/o scenario-generator & 0 & 0 & 0 & 0 & 0 \\
        w/o reflection-loop & 0 & 0 & 0 & 0 & 0 \\
        \bottomrule
    \end{tabular}
    }
\end{table}

\subsection{Communication Mechanism}

Communication in \consensustester involves the memory, the global state, and the consensus protocol code repository. 

As discussed in \S\ref{sec:memory}, the memory is lightweight and is accessed by the agents only once during the entire execution. By maintaining the global state within \agenta, \agentb does not need to track previously generated attack scenarios or discovered bugs and can easily retrieve the required state information from \agenta. Similarly, \agentc obtains the attack scenarios through \agenta. As a result, \agentb and \agentc do not need to manage contextual information and can focus solely on their respective tasks, while \agenta maintains the system-wide state and effectively manages long-horizon objectives.
The global state itself is succinct, and its transmission occurs only between \agenta and \agentb or \agentc, without direct communication between \agentb and \agentc. Such communication is necessary to maintain system consistency and cannot be further reduced. In \consensustester, agents can access the consensus protocol code repositories via the Model Context Protocol (MCP)~\cite{anthropic2024mcp}.

Overall, this design keeps the system concise, with communication limited to essential information, resulting in low communication overhead.

\section{Experiments}
\label{sec:exp}

\subsection{Experiment Setup}

\textbf{Baselines.}
Existing studies that leverage LLMs for bug detection either lack integration with multi-agent systems or do not focus on deep logic-level bugs~\cite{Kaniewski2025ASL,Sheng2025LLMsIS}. To address these limitations and obtain appropriate comparison baselines, we adapt ReAct~\cite{yao2022react} by augmenting it with code-reading tools, and implement it using state-of-the-art code-oriented large language models as baselines.
Specifically, we evaluate three commercial LLMs, Claude Sonnet 4.5~\cite{anthropic2025claude45}, Gemini 3.0 Pro Preview~\cite{google2025gemini3}, and GPT-5.2~\cite{openai2025gpt52}, as well as one open-source model, Qwen3 Coder 480B A35B~\cite{qwen2025qwen3}.

\begin{figure}[!t]
  \centering
  \includegraphics[width=0.8\linewidth]{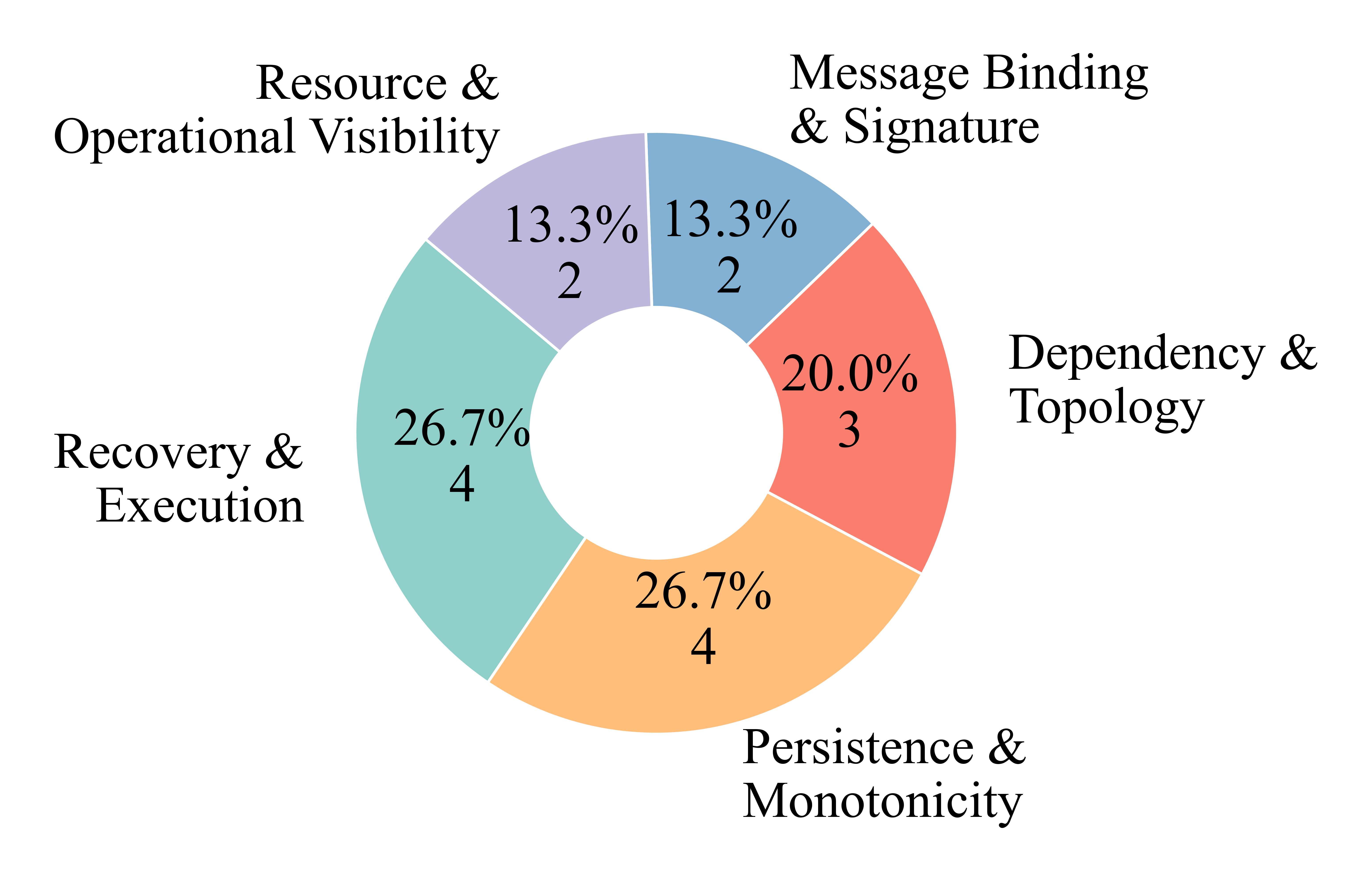}
  \caption{Distribution of discovered 15 protocol-level logic bugs of \consensustester by category.}
  \label{fig:bugclass}
\end{figure}

\textbf{Implementation Details.}
For our implementation, we access all models via APIs, eliminating the need for GPUs. For all large language models, we set the temperature to 0.2. In each experiment, both \consensustester and the baselines are allowed to run for three hours. For \consensustester, the number of reflection loop iterations (\texttt{num\_iter}) is set to 50. The implementation and artifact details are in Appendix~\ref{appendsec:artifact}.

\begin{figure*}[!t]
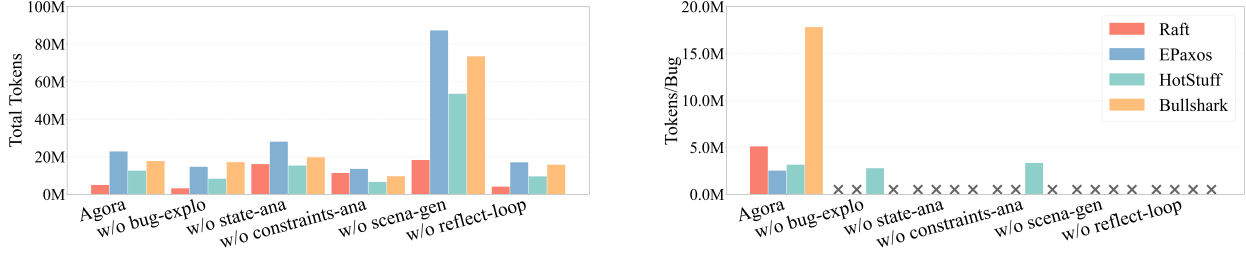

    \centering
    \subfloat{
        \includegraphics[width=0.45\linewidth]{pic/total_tokens.png}
    }
     \hspace{0.04\linewidth}
    \subfloat{
    \includegraphics[width=0.45\linewidth]{pic/tokens_per_bug.png}
    }
    \caption{The left (a) shows the total token consumption in the ablation studies.
The right (b) shows the average token consumption for each discovered bug in the ablation studies, where ``cross" indicates that no bugs are discovered.}
\label{fig:tokens}
\end{figure*}

\textbf{Code Repositories.} We evaluate our approach on four popular protocol code repositories, covering two BFT and two CFT protocols, including Raft (etcd~\cite{etcdioraft}), EPaxos (efficientEPaxos~\cite{efficientepaxos}), HotStuff (relabHotStuff~\cite{relabhotstuff}), and BullShark (sui~\cite{suimystenlabs}). Among these, etcd and sui are production-level code repositories. Details of these repositories are in Appendix~\ref{appendsec:codebase}.

\subsection{Overall Results}
\label{sec:expoverall}

Table~\ref{tab:overall} presents the overall experimental results of \consensustester compared with the baselines. From the results, we observe that regardless of the underlying LLM, the baselines are only able to identify simple implementation-level bugs, whereas our \consensustester, \textbf{by design}, consistently discovers protocol-level logic bugs and does not report implementation bugs.

Implementation-level bugs can typically be detected through static analysis and often result in issues such as integer overflows or memory violations. In contrast, logic bugs are significantly more subtle and may compromise system correctness, data consistency, or transaction security, leading to severe real-world consequences. Therefore, \consensustester explicitly focuses on deep protocol logic bugs rather than implementation-level issues. Detailed examples of the two bug types are provided in Appendix~\ref{appendsec:bugexample}. The experimental results demonstrate that our multi-agent design enables \consensustester to effectively focus on the detection of deep logic bugs, saving the computational resources and validating the effectiveness of our approach.

More specifically, we find that even when augmented with ReAct design, baselines using Gemini 3.0 Pro Preview and Claude Sonnet 4.5 fail to identify any implementation-level bugs. GPT-5.2 is the most effective baseline model, detecting 22 bugs across three repositories. Qwen3 Coder 480B A35B, possibly benefiting from its optimization for code understanding, is able to detect three bugs in the EPaxos repository, all of which are also identified by GPT-5.2. In total, the baselines uncover 22 zero-day implementation bugs, which we report as an additional contribution toward improving robustness of existing consensus implementations.

In contrast, \consensustester achieves the best performance when built upon Gemini 3.0 Pro Preview, identifying 11 protocol-level logic bugs across four widely used repositories, including production-level systems. This result is particularly notable given that the Gemini-based baseline fails to detect any bugs. Moreover, \consensustester consistently discovers logic bugs when instantiated with each of the other three LLMs, further demonstrating the robustness and generality of our approach. As some bugs are discovered by multiple LLM-backed instances, \consensustester identifies a total of 15 unique zero-day logic bugs, substantially enhancing the security of distributed consensus systems. These results further validate the effectiveness of integrating our framework with the HDT paradigm.

\textbf{Practical Usage.} We further analyze all the 46 bug reports generated by the \agentc agent in \consensustester across the four LLMs. Among them, 34 reports correspond to real protocol-level logic bugs, resulting in a false positive rate of \textbf{only 26.1\%}. This low false positive rate indicates that \consensustester does not require extensive manual filtering, making it a highly efficient and practical tool for engineers and researchers to discover deep logic bugs in real-world systems.

\textbf{Bug Summary.}
Based on the categories of consensus logic bugs, we further analyze the 15 bugs discovered by \consensustester. As shown in Figure~\ref{fig:bugclass}, violations are observed across all bug categories, demonstrating that \consensustester is capable of triggering a wide range of deep protocol-level bugs.
Specifically, Recovery \& Execution Divergence bugs account for four cases (26.7\%), tying with Persistence \& Monotonicity Violations as the most prevalent category. In addition, Message Binding \& Signature Violations, Resource \& Operational Visibility Violations, and Dependency \& Topology Flaws account for 2, 2, and 3 bugs, respectively. Detailed examples of logic bugs from different categories are in Appendix~\ref{appendsec:bugclassexample}.

\subsection{Ablation Studies}\label{sec:ablation}

In this subsection, we analyze the impact of the five components of \consensustester on both bug discovery effectiveness and the computational resources required to discover bugs. Following the same experimental settings as before, we conduct ablation studies and select Gemini 3.0 Pro Preview as the underlying LLM, as it achieves the best overall performance under overall experiments in \S\ref{sec:expoverall}.

\textbf{Summary.} Table~\ref{tab:ablation_study} summarizes the bug discovery results in the ablation study. The results show that only the complete \consensustester is able to discover the largest number of bugs (11 bugs), while removing any single component leads to degraded performance (at most 3 bugs). Overall, ablating any component reduces effectiveness by \textbf{73–100\%}.
Figure~\ref{fig:tokens}a shows the total token consumption in our ablation studies. Although removing the \texttt{bug-exploitation} and \texttt{reflection-loop} components can slightly reduce token usage (43.8M and 46.9M tokens in total, respectively, compared to 58.6M for the original configuration), this significantly hampers the discovery of useful bugs. Removing any of the other components results in higher token consumption, with the most significant increase observed when removing the \texttt{scenario-generator} (we will discuss later). Combined with Table~\ref{tab:ablation_study}, these results demonstrate that our modular design enables \consensustester to discover more meaningful bugs with fewer tokens, validating the effectiveness of our proposed components. Figure~\ref{fig:tokens}b further reports the number of tokens required to discover each bug under different ablation settings. Since majority of experiments fail to discover any bugs, \consensustester is the most efficient, discovering one critical logic bug with an average of 5.32M tokens (approximately \$40). Identifying such protocol-level logic bugs typically requires years of effort from experienced engineers, making \consensustester highly cost-effective in practice.

\textbf{Bug-exploitation.} Removing bug-exploitation component is still able to discover a small number of bugs (3 bugs) in HotStuff, but fails to identify logic bugs in other repositories. Moreover, it cannot further explore deeper bugs based on previously discovered ones, resulting in significant token waste. Detailed examples of deeper bugs triggered through bug exploitation are provided in Appendix~\ref{appendsec:deepbug}. These results demonstrate that bug-exploitation enables \consensustester to more effectively discover deeper protocol-level bugs.

\textbf{State-analyzer.} Without state-analyzer component, \consensustester fails to discover any bugs. Further analysis shows that the system loses essential execution context, repeatedly retries similar scenarios, consumes excessive tokens (79.6M), and is unable to uncover logic bugs.

\textbf{Constraints-analyzer.} Removing constraints-analyzer component allows the system to discover a single bug in HotStuff (BFT protocol), but leads to a large number of false positives when testing CFT systems. In our experiments, this configuration generates \textbf{107} CFT bug reports, none of which correspond to positive logic bugs after manual inspection. This highlights the necessity of integrating the constraints-analyzer into \consensustester. Detailed differences between BFT and CFT bugs are discussed in Appendix~\ref{appendsec:bftbug} and Appendix~\ref{appendsec:cftbug}.

\textbf{Scenario-generator.}  When the scenario-generator component is removed, \consensustester regresses to baseline behavior. As shown in Figure~\ref{fig:tokens}a, this configuration incurs the highest token consumption (233.1M tokens, \textbf{3.98} times of original), discovers numerous implementation-level bugs, yet fails to identify any protocol-level logic bugs. In this setting, we uncover \textbf{29} zero-day implementation-level bugs. Although this demonstrates the effectiveness of the remaining components in enabling iterative bug discovery, it wastes tokens and fails to discover deep bugs targeted by our design goal.

\textbf{Reflection-loop.} Removing reflection-loop component prevents \consensustester from generating correct unit test files, thereby breaking the iterative discovery process for logic bugs. This results in wasted tokens, human power and reduced efficiency due to the generation of \textbf{173} problematic bug reports.

In summary, \agenta agent leverages the bug-exploitation and state-analyzer component to reduce context usage and communication overhead; \agentb agent generates effective attack scenarios for discovering deep bugs and adapts to different protocol types through the constraints-analyzer; and \agentc agent relies on the reflection-loop component to generate correct test cases. Together, these components ensure that \consensustester achieves the key properties of interpretability, efficiency, completeness, and flexibility.

\section{Limitations and Discussions}
\label{sec:limitation}

Although our system has demonstrated significant effectiveness and has uncovered many zero-day logic bugs in consensus implementations, it still requires a certain amount of human knowledge. In the future, we aim to further improve its level of automation.
Due to the inherent difficulty of identifying consensus bugs, we find 15 bug instances. In future work, we plan to extend our bug exploitation techniques and deploy our experiments on a broader range of consensus protocol codebases to collect sufficient data. With more data, we intend to apply Supervised Fine-Tuning~\cite{wei2022finetuned,Ouyang2022TrainingLM} to enhance our system, and further generate trajectories to finetune our agent system using Direct Preference Optimization~\cite{Rafailov2023DirectPO}.

Moreover, we plan to generalize \consensustester, leveraging HDT and our decoupled design, to other debugging domains where both protocols and implementations are highly coupled and complex, such as concurrency control~\cite{Tan2020CobraMT,Clark2024ValidatingDS} and smart contracts~\cite{Sun2024GPTScanDL}, which are prone to causing severe economic and security issues in real-world systems.

\section{Conclusion}
\label{sec:conclusion}

In this paper, we combine specific domain knowledge with an LLM-based multi-agent system and, following the HDT paradigm, propose the \consensustester framework to detect consensus protocol–level logic bugs. Our approach is applicable to both research prototypes and production-level code repositories. By generating targeted attack scenarios, \consensustester is able to uncover deep, high-impact vulnerabilities that pose serious real-world threats. The system supports multiple programming languages, significantly enhancing the security of distributed systems, and demonstrates clear advantages over existing baselines.
Benefiting from our cleanly decoupled design, our multi-agent framework not only satisfies key properties such as effectiveness, interpretability, efficiency, completeness, and flexibility, but also provides a strong foundation for future extension to other domain knowledge–intensive bug detection tasks.

\section*{Impact Statement}
This paper presents work whose goal is to advance the field of Machine Learning. There may be some potential societal consequences of our work, none of which we feel must be specifically highlighted here.



\bibliography{example_paper}
\bibliographystyle{icml2026}

\newpage
\appendix
\onecolumn

\begin{center}
    {\LARGE \bf Appendix}
\end{center}
\begin{quote}
    \hypersetup{linkcolor=black} 
    \textbf{Contents}
    \begin{itemize}
    \item \hyperref[appendsec:artifact]{Section \ref{appendsec:artifact}: Artifact Details}
        \item \hyperref[appendsec:multiagent]{Section \ref{appendsec:multiagent}: Mathematical Modeling of the Multi-Agent System}
        \item \hyperref[appendsec:bugpattern]{Section \ref{appendsec:bugpattern}: Bug Pattern Details}
        \item \hyperref[appendsec:constraints]{Section \ref{appendsec:constraints}: Constraints Details}
        \item \hyperref[appendsec:memory]{Section \ref{appendsec:memory}: Memory Details}
        \item \hyperref[appendsec:codebase]{Section \ref{appendsec:codebase}: Codebase Details}
        \item \hyperref[appendsec:bugexample]{Section \ref{appendsec:bugexample}: Two Types of Bug Example}
        \item \hyperref[appendsec:bugclassexample]{Section \ref{appendsec:bugclassexample}: Bug Example for Five Classes}
        \item \hyperref[appendsec:prompt]{Section \ref{appendsec:prompt}: Prompt Example}
    \end{itemize}
\end{quote}
\hrule


\section{Artifact Details}
\label{appendsec:artifact}
Our codebase is available at \url{https://github.com/lebronlambert/Agora}. Similar to \consensustester, our baselines also incorporate essential information, such as repository knowledge to get a fair comparison. Setting up the environment is straightforward with the provided README file. In the README, you will find all the necessary guidelines to use the bash scripts to reproduce the main results from our experiments.

\section{Mathematical Modeling of the Multi-Agent System}
\label{appendsec:multiagent}
LLMs power a wide variety of applications~\cite{
cao2026taskspecificefficiencyanalysissmall,zeng2025bridging}. Specifically, in a formal sense, we characterize an autonomous LLM-powered entity as a quintuple $\mathcal{E} = \langle \mathcal{B}, \mathcal{P}, \mathcal{A}, \Omega, \Gamma \rangle$. Here, $\mathcal{B}$ represents the core linguistic brain (the LLM), $\mathcal{P}$ signifies the perceptual manifold of the target environment, and $\mathcal{A}$ denotes the repertoire of executable operations (tools). The mapping function $\Omega: \mathcal{P} \to \mathcal{A}$ acts as the decision logic, while $\Gamma$ encapsulates the persistent memory state. The operational trajectory of such an entity follows a recursive observe-deliberate-act cycle: at any discrete temporal index $k$, the entity ingests a sensory input $p_k \in \mathcal{P}$, leverages $\mathcal{B}$ to synthesize a strategic intent based on $\Omega$, and dispatches an intervention $a_k \in \mathcal{A}$ to modulate the system state. To tackle the intricate state spaces of consensus implementations, we extend this individual paradigm into a collaborative collective $\mathcal{G} = \{ \mathcal{E}_1, \mathcal{E}_2, \dots, \mathcal{E}_m, \mathcal{X} \}$. In this formulation, each $\mathcal{E}_i$ represents a specialized functional unit, and $\mathcal{X}$ defines the coordination medium that facilitates information exchange and role synchronization. This multi-agent architecture transcends the cognitive boundaries of isolated models by decomposing complex protocol verification into modular, inter-dependent reasoning sub-tasks, a methodology that has proven more robust than monolithic approaches in recent literature, e.g., CAMEL~\cite{Li2023CAMELCA}, MetaGPT~\cite{hong2023metagpt}, OWL~\cite{Hu2025OWLOW}.

\section{Bug Pattern Details}
\label{appendsec:bugpattern}
We implemented a crawler to collect all \textbf{816} GitHub issues from the four code repositories. We then filtered the issues and identified \textbf{34} issues related to consensus protocol logic bugs, which were used to construct our bug patterns to help us integrate expert knowledge. Examples are shown below:

\PatternBox{Bug Pattern Details Show Case 1}
{
\textbf{Pattern ID:} Byzantine DoS via insufficient QC validation - single malicious node can stall consensus by forging high-view QCs \\
\textbf{Protocol Type:} BFT (HotStuff Variant) \\
\textbf{Category:} Security / Liveness / Insufficient Validation

\vspace{0.5em}
\textbf{Description:}
A critical vulnerability exists in the \texttt{AggregateQC} creation and verification process. A single Byzantine node can indefinitely stall consensus by providing a forged Quorum Certificate (QC) with a deceptively high view number.

\vspace{0.5em}
\textbf{Root Cause Analysis:}
\begin{itemize}
    \item \texttt{CreateAggregateQC} blindly trusts and collects QCs from \texttt{TimeoutMsg} payloads without individual verification.
    \item \texttt{VerifyAggregateQC} only validates the \texttt{highQC} (the QC with the highest view), creating a single point of failure.
\end{itemize}

\vspace{0.5em}
\textbf{Attack Scenario:}
\begin{enumerate}
    \item \textbf{Forgery:} A Byzantine node crafts a \texttt{TimeoutMsg} containing a forged QC with a very high view number.
    \item \textbf{Aggregation:} The Leader trustingly includes this invalid \texttt{highQC} into the \texttt{AggregateQC}.
    \item \textbf{Proposal:} The Leader proposes the new block/view containing this \texttt{AggregateQC}.
    \item \textbf{Rejection:} Honest nodes attempt to verify the \texttt{AggregateQC}. Since \texttt{VerifyQuorumCert(highQC)} fails, the entire proposal is rejected.
    \item \textbf{Stall:} As the leader cannot form a valid proposal that honest nodes accept, the consensus process halts.
\end{enumerate}

\vspace{0.5em}
\textbf{Impact:}
Complete blockchain liveness failure. A single malicious actor can perpetually halt the network's progress.

\vspace{0.5em}
\textbf{Fix \& Mitigation:}
\begin{itemize}
    \item \textbf{Individual Verification:} Modify \texttt{CreateAggregateQC} to independently validate every QC using \texttt{VerifyQuorumCert} before its inclusion.
    \item \textbf{Zero-Trust Policy:} Never trust aggregate structures from other nodes without verifying each component.
\end{itemize}

\vspace{0.5em}
\textbf{Context:}
\begin{itemize}
    \item \textbf{Reference:} \href{https://github.com/relab/hotstuff/issues/224}{relab/hotstuff Issue \#224}
    \item \textbf{Vulnerable Functions:} \texttt{CreateAggregateQC}, \texttt{VerifyAggregateQC}
    \item \textbf{Confirmed by:} meling (maintainer)
\end{itemize}
}

\PatternBox{Bug Pattern Details Show Case 2}
{
\textbf{Pattern ID:} VerifyLeader OOM - notify list grows unbounded when futures not removed from repl.notify on error \\
\textbf{Protocol Type:} CFT (Hashicorp Raft) \\
\textbf{Bug Category:} Liveness / Memory Leak / OOM

\vspace{0.5em}
\textbf{Description:}
In the \texttt{VerifyLeader} mechanism, a node failure triggers an unbounded memory growth. The \texttt{verifyLeader()} function registers a verify future \texttt{v} into the \texttt{repl.notify} list for each replica. When a heartbeat or \texttt{AppendEntries} returns an error, the cleanup process is incomplete, leading to a memory leak that eventually causes an Out-Of-Memory (OOM) crash.

\vspace{0.5em}
\textbf{Root Cause Analysis:}
\begin{itemize}
    \item \textbf{Asymmetric Cleanup:} The future \texttt{v} is tracked in two distinct structures: \texttt{leaderState.notify} and \texttt{repl.notify}.
    \item \textbf{Incomplete Removal:} On error, the future is only deleted from \texttt{leaderState.notify} but remains in \texttt{repl.notify}.
    \item \textbf{Resource Exhaustion:} The replication goroutine persists in holding references to failed or completed futures, preventing Garbage Collection (GC) and causing the list to grow infinitely.
\end{itemize}

\vspace{0.5em}
\textbf{Lessons \& Best Practices:}
\begin{itemize}
    \item \textbf{Symmetric Lifecycle:} Ensure that any future or callback registered in multiple structures is removed from \textit{all} of them upon completion or failure.
    \item \textbf{Failure Path Testing:} Monitor memory usage specifically under failure conditions, as leaks often remain hidden during healthy cluster operations.
    \item \textbf{Cleanup Patterns:} Utilize \texttt{defer} or RAII-like patterns to guarantee cleanup across all possible execution paths.
    \item \textbf{Subscription Management:} For async notifications, consider explicit lifecycle management or weak references to manage subscription longevity.
\end{itemize}

\vspace{0.5em}
\textbf{Context:}
\begin{itemize}
    \item \textbf{Reference:} \href{https://github.com/hashicorp/raft/issues/291}{Hashicorp Raft Issue \#291}
    \item \textbf{Tags:} \texttt{cft}, \texttt{memory\_leak}, \texttt{verify\_leader}, \texttt{async\_notify}
\end{itemize}
}

\section{Constraints Details}
\label{appendsec:constraints}

Here, we provide examples of constraints generated for BFT and CFT systems, which help generate scenarios tailored to different types of consensus protocols.

\PatternBox{Protocol Constraints \& BFT Attack Model}
{
\textbf{Protocol Type Constraint:} BFT (Byzantine Fault Tolerant) \\
This model assumes an environment capable of tolerating \textbf{Byzantine faults}, where nodes may behave maliciously or arbitrarily.

\vspace{0.8em}
\textbf{I. CFT-type Attack Vectors (Crash Faults):}
\begin{itemize}
    \item \textbf{Availability:} Node crash, recovery, and frequent restarts.
    \item \textbf{Network:} Partitions, significant latency, packet loss, and message reordering.
    \item \textbf{Logic:} Boundary condition exploits and duplicate message utilization.
\end{itemize}

\vspace{0.5em}
\textbf{II. BFT-specific Attack Vectors (Malicious Behavior):}
\begin{itemize}
    \item \textbf{Equivocation:} Sending conflicting votes/proposals for the same round/height to different peers.
    \item \textbf{Payload Tampering:} Injecting invalid, malformed, or tampered protocol messages.
    \item \textbf{Selective Broadcast:} Distributing messages to a subset of nodes to induce state inconsistency.
    \item \textbf{Strategic Withholding:} A malicious leader intentionally delays or withholds proposals to stall liveness.
    \item \textbf{State Forgery:} Crafting fraudulent certificates, votes, or cryptographic signatures.
    \item \textbf{Timing Attacks:} Exploiting vulnerabilities within the protocol's timing and timeout assumptions.
\end{itemize}

\vspace{0.8em}
\textbf{III. Attack Mindset \& Objectives:}
\textit{Assuming control of up to $f$ malicious nodes ($f < n/3$), the attacker aims to:}
\begin{enumerate}
    \item \textbf{Safety Violation:} Can honest nodes be coerced into accepting an invalid state?
    \item \textbf{Liveness Violation:} Can the system be prevented from reaching consensus indefinitely?
    \item \textbf{Consistency Violation:} Can different honest nodes be led to perceive divergent ``truths''?
\end{enumerate}

\vspace{0.5em}
\textbf{Note:} Effective testing must simulate a clever attacker who strategically exploits the protocol's specific logic weaknesses rather than just random fault injection.
}

\PatternBox{Protocol Constraints \& CFT Fault Model}
{
\textbf{Protocol Type Constraint:} CFT (Crash Fault Tolerant) \\
This model assumes an environment that can only tolerate \textbf{crash faults}. All nodes are assumed to be honest; they either follow the protocol correctly or stop functioning entirely.

\vspace{0.8em}
\textbf{I. Allowed Attack Methods (Non-Malicious Faults):}
\begin{itemize}
    \item \textbf{Node Dynamics:} Random node crashes (immediate stop) and restarts (recovery with potential loss of in-memory state).
    \item \textbf{Network Adversity:} Network partitions, extreme communication delays, message loss, and packet reordering.
    \item \textbf{Implementation Limits:} Boundary conditions (overflows, null values, extreme inputs) and duplicate resource utilization.
\end{itemize}

\vspace{0.5em}
\textbf{II. Prohibited Attack Methods (Byzantine Behavior):}
\begin{itemize}
    \item \textbf{No Malicious Messages:} Sending forged, tampered, or intentionally malformed payloads is \textit{strictly prohibited}.
    \item \textbf{No Equivocation:} Double voting or sending conflicting proposals in the same round is not allowed.
    \item \textbf{No Selective Broadcast:} Intentional inconsistency through partial message distribution is excluded.
    \item \textbf{No Identity Theft:} Actions involving signature forgery or impersonating other nodes are disallowed.
\end{itemize}

\vspace{0.8em}
\textbf{III. Core Assumptions:}
\textit{In a CFT environment, the testing focus shifts from "malice" to "reliability":}
\begin{enumerate}
    \item \textbf{Binary State:} Nodes are either "Correct" (following the protocol) or "Failed" (stopped). There is no "half-dead" or Byzantine behavior.
    \item \textbf{Honest Participation:} All participating nodes are assumed to act in good faith to reach a common consensus.
    \item \textbf{Recovery Consistency:} The system must maintain safety and eventually regain liveness after network partitions heal or nodes recover.
\end{enumerate}

\vspace{0.5em}
\textbf{Note:} Testing for CFT protocols should exhaustively simulate unstable infrastructure and edge-case logic rather than adversarial intent.
}

\section{Memory Details}
\label{appendsec:memory}


The memory contains repository knowledge and bug patterns, with a small total size of about 360 KB. Examples of repository knowledge are shown below.

\PatternBox{Repository Knowledge: etcd Raft (Go)}
{
\textbf{Core Testing Helpers \& API:}
\begin{itemize}
    \item \texttt{newNetwork(peers...)}: Creates an in-memory Raft network simulation environment, supporting fault injection and message dropping.
    \item \texttt{newTestRaft(id, election, heartbeat, storage)}: Initializes a Raft instance for white-box testing with specific timing parameters.
    \item \texttt{nextEnts(r, s)}: Returns entries that are committed but not yet applied, and advances the \texttt{applied} index.
    \item \texttt{MemoryStorage}: In-memory implementation of the \texttt{Storage} interface. Key operations like \texttt{Append}, \texttt{CreateSnapshot}, and \texttt{Compact} must respect boundary conditions (e.g., \texttt{Compact} cannot exceed \texttt{lastIndex}).
\end{itemize}

\vspace{0.5em}
\textbf{Key States \& Message Types:}
\begin{itemize}
    \item \textbf{States:} \texttt{StateFollower} (0), \texttt{StateCandidate} (1), \texttt{StateLeader} (2), \texttt{StatePreCandidate} (3).
    \item \textbf{Local Triggers:} \texttt{MsgHup} (triggers election), \texttt{MsgBeat} (triggers heartbeat broadcast).
    \item \textbf{Replication:} \texttt{MsgProp} (proposal), \texttt{MsgApp} (append entries), \texttt{MsgAppResp} (append response).
    \item \textbf{Control:} \texttt{MsgHeartbeat}, \texttt{MsgSnap} (snapshot transfer), \texttt{MsgVote} (solicit vote).
\end{itemize}

\vspace{0.8em}
\textbf{Lessons Learned \& Common Pitfalls:}
\begin{itemize}
    \item \textbf{Implicit No-op Entry:} Upon becoming a Leader (\texttt{becomeLeader}), the node automatically appends an empty \textbf{no-op} entry. This must be accounted for when calculating expected \texttt{lastIndex}.
    \item \textbf{Single-Node Quorum:} In a 1-node cluster, entries are committed immediately upon append, but still require a call to \texttt{nextEnts} to drive the application logic.
    \item \textbf{Flow Control (Inflights):} When \texttt{MaxInflight=1}, the system is highly sensitive. Probe responses (empty \texttt{MsgApp}) must be handled correctly to unpause the pipeline and prevent \texttt{PipelineStall}.
    \item \textbf{Snapshot Boundaries:} \texttt{MemoryStorage} maintains a dummy entry at \texttt{FirstIndex()-1}. If \texttt{HardState.Commit} is less than \texttt{FirstIndex}, Raft will panic during initialization.
    \item \textbf{Learner Promotion Safety:} Promoting a Learner to a Voter while it has unacknowledged inflight messages can lead to state machine stalls (e.g., stuck in \texttt{StateProbe}) if the transition logic does not explicitly clear buffers.
\end{itemize}

\vspace{0.5em}
\textbf{Execution Context:} 
Always run tests using the package context (\texttt{go test -v -run TestName .}) rather than standalone files (\texttt{go test file.go}) to ensure that unexported helper functions in \texttt{raft\_test.go} are correctly linked.
}

\begin{table}[!t]
\caption{Codebase information.}
\vspace{0.5em}
\label{tab:codebase}
\centering
\resizebox{0.7\textwidth}{!}{
\begin{tabular}{l c l l r r }
\toprule
\textbf{Repo} & \textbf{Type} & \textbf{Protocol} & \textbf{Lang} & \textbf{Stars} & \textbf{Forks} \\
\midrule
etcd~\cite{etcdioraft} & CFT & Raft & Go & 982 & 222 \\
efficientEPaxos~\cite{efficientepaxos} & CFT & EPaxos & Go & 631 & 138 \\
relabHotStuff~\cite{relabhotstuff} & BFT & HotStuff & Go & 217 & 67 \\
sui~\cite{suimystenlabs} & BFT & BullShark & Rust & 7.6k & 11.7k \\
\bottomrule
\end{tabular}
}
\end{table}

\section{Codebase Details}
\label{appendsec:codebase}
We evaluate our approach on four popular consensus protocol codebases, including both research prototypes and production-level repositories, as summarized in Table~\ref{tab:codebase}.

\section{Two Types of Bug Example}
\label{appendsec:bugexample}
\subsection{Implementation Bug}
\label{appendsec:implementationbug}

\BugAnalysisBox{Implementation Bug}
{
\textbf{Incorrect RPC Message Dispatch (Copy-Paste Error)}

\paragraph{Problem Description}
A critical logic error exists in the \texttt{bcastTryPreAccept} function. Despite correctly populating a \texttt{TryPreAccept} structure (\texttt{tpa}), the code erroneously references an unrelated \texttt{PreAccept} structure (\texttt{pa}) during dispatch.

\paragraph{Affected Components}
\begin{itemize}
    \item \textbf{File:} \texttt{src/epaxos/epaxos.go}
    \item \textbf{Function:} \texttt{bcastTryPreAccept} (Lines 580--595)
\end{itemize}

\paragraph{Root Cause Analysis}
In \texttt{epaxos.go}, the function spends lines 588--594 filling the fields of \texttt{tpa}. However, at line 595, the code executes: \texttt{args := \&pa}. Since \texttt{pa} represents a different RPC schema, the broadcast sends stale or misaligned protocol data.

\paragraph{Attack Scenario / Impact}
\begin{enumerate}
    \item \textbf{Mangled Communication:} The leader broadcasts a message intended as \texttt{TryPreAccept}, but the payload is cast from a \texttt{PreAccept} structure.
    \item \textbf{Protocol Collapse:} 
        \begin{itemize}
            \item \textbf{Recovery Failure:} Doubtful instances remain unresolved, leading to a permanent Liveness hang.
            \item \textbf{State Machine Divergence:} Peers may accept inconsistent dependency graphs based on outdated \texttt{pa} data.
        \end{itemize}
\end{enumerate}

\paragraph{Corrective Action}
Update the reference at line 595 to the correct initialized structure:
\begin{center}
    \texttt{Change \textbf{\&pa} $\rightarrow$ \textbf{\&tpa}}
\end{center}
} 

\subsection{Logic Bug}
\label{appendsec:logicbug}

\subsubsection{BFT Bug}
\label{appendsec:bftbug}
\BugAnalysisBox{BFT Bug}
{
\textbf{Case Study: Safety Violation via Cascading Forks in HotStuff}

\paragraph{Affected Components}
\begin{itemize}
    \item \textbf{Files:} \texttt{protocol/rules/chainedhotstuff.go}, \texttt{protocol/consensus/voter.go}
    \item \textbf{Modules:} \texttt{ChainedHotStuff.VoteRule()}, \texttt{CommitRule()}
\end{itemize}

\paragraph{Root Cause Analysis}
The vulnerability stems from an interplay between weak liveness checks and the 3-chain commitment logic:
\begin{itemize}
    \item \textbf{Vulnerable Vote Rule:} The liveness condition (\texttt{QC.View > bLock.View}) is too permissive. If a node's \texttt{bLock} is lagged (e.g., after a reboot), any Quorum Certificate (QC) from a higher view can bypass the safety check, even if it extends an ancient branch.
    \item \textbf{Context-Free Commit Rule:} The \texttt{CommitRule} strictly validates the 3-chain structure (consecutive views and matching parent hashes) but fails to ensure the chain is part of the "canonical" branch.
    \item \textbf{Mutable bLock:} \texttt{bLock} updates whenever a 2-chain is formed on \textit{any} fork, lowering the barrier for Byzantine leaders to push a malicious branch.
\end{itemize}

\paragraph{Attack Scenario (n=4, f=1)}
\begin{enumerate}
    \item \textbf{State:} Honest nodes hold a high QC for $B_3$ (Chain: $G \rightarrow B_1 \rightarrow B_2 \rightarrow B_3$). A target honest node has \texttt{bLock = Genesis}.
    \item \textbf{Step 1 (View 4):} A Byzantine leader proposes $B_4'$ extending $B_1$. Since $QC_{B1}.View (1) > bLock.View (0)$, the node votes, forming $QC_4'$.
    \item \textbf{Step 2 (Views 5--6):} Consecutive Byzantine leaders extend the fork to $B_5'$ and $B_6'$. The node's \texttt{bLock} is updated to $B_4'$, facilitating the fork's progression.
    \item \textbf{Step 3 (View 7):} Upon receiving $B_7'$, the \texttt{CommitRule} detects a valid 3-chain ($B_4' \rightarrow B_5' \rightarrow B_6'$). Consequently, $B_4'$ is committed, bypassing the correct $B_2 \rightarrow B_3$ sequence.
\end{enumerate}

\paragraph{Impact \& Security Analysis}
This attack is critical as it leads to a \textbf{Safety Violation}. Unlike transient liveness issues, this results in permanent state pollution and ledger divergence across honest nodes. The attack is feasible in scenarios involving node restarts or state synchronization lags, especially when the protocol allows a Byzantine node to be elected as a leader for consecutive views.

\paragraph{Verification}
The exploit was successfully reproduced using:\\
\texttt{go test -v -run TestByzantineLeaderBuildsCascadingForkAcrossViews}
}

\subsubsection{CFT Bug}
\label{appendsec:cftbug}

\BugAnalysisBox{CFT Bug}
{
\textbf{Case Study: Dependency Divergence during EPaxos Recovery}

\paragraph{Affected Components}
\begin{itemize}
    \item \textbf{File:} \texttt{src/epaxos/epaxos.go}
    \item \textbf{Modules:} \texttt{handlePrepareReply()}, \texttt{startPhase1()}, and \texttt{updateAttributes()}
\end{itemize}

\paragraph{Root Cause Analysis}
The vulnerability is rooted in the improper re-computation of command attributes during the recovery phase. When a replica attempts to recover an instance whose Fast Path status is ambiguous, it incorrectly invokes \texttt{startPhase1}, which triggers a fresh call to \texttt{updateAttributes}.
\begin{itemize}
    \item \textbf{Attribute Volatility:} Instead of preserving the original dependencies recorded during the pre-accept phase, \texttt{startPhase1} re-scans the local log for conflicting commands.
    \item \textbf{Dependency Smuggling:} If a new, conflicting command has arrived during the leader's downtime, it will be "smuggled" into the recovered command's dependency set. This creates a discrepancy between the dependencies used during the original Fast Path execution and the final committed state.
\end{itemize}

\paragraph{Attack Scenario (3 Nodes: $R_0, R_1, R_2$)}
\begin{enumerate}
    \item \textbf{Fast Path Execution:} $R_0$ proposes $Cmd_A$ (Key="x"). $R_0$ and $R_1$ accept it with $\text{deps}=\{\}$. $R_0$ reaches a Fast Quorum, commits, and executes $Cmd_A$ locally ($x=1$).
    \item \textbf{Leader Crash:} $R_0$ crashes immediately before broadcasting the \texttt{Commit} message to others.
    \item \textbf{Conflict Injection:} $R_2$ proposes $Cmd_B$ (conflicting with $Cmd_A$). $R_2$ and $R_1$ accept $Cmd_B$. Crucially, $R_1$ now sees $Cmd_B$ in its log.
    \item \textbf{Vulnerability Trigger:} $R_1$ initiates recovery for $Cmd_A$. Failing to observe a Fast Quorum, it executes \texttt{startPhase1}. The \texttt{updateAttributes} function sees $Cmd_B$ and re-calculates $Cmd_A$'s dependencies as $\text{deps}=\{Cmd_B\}$.
    \item \textbf{State Divergence:} The cluster eventually commits $Cmd_A$ with $\text{deps}=\{Cmd_B\}$, implying an execution order of $B \rightarrow A$. However, the failed leader $R_0$ has already executed $A$ with $\text{deps}=\{\}$.
\end{enumerate}

\paragraph{Impact \& Security Analysis}
This bug represents a \textbf{Linearizability Violation}. Upon recovery, $R_0$ will find its local execution history contradicts the globally committed serialized order. In high-throughput systems with frequent leader transitions or network partitions, this divergence leads to permanent database corruption and "split-brain" states at the application level.

\paragraph{Verification}
The divergence can be reproduced by a controlled test case where:
\begin{itemize}
    \item A specific node is isolated after a Fast Path commit.
    \item A conflicting command is injected into the remaining quorum.
    \item Recovery is forced on the surviving nodes.
\end{itemize}
}

\subsubsection{Bug Triggers from Bug Exploitation}
\label{appendsec:deepbug}
\BugAnalysisBox{Bug from Bug Exploitation}
{
\textbf{Case Study: History Erasure via Deep Fork from Genesis}

\paragraph{Affected Components}
\begin{itemize}
    \item \textbf{Files:} \texttt{protocol/rules/chainedhotstuff.go}, \texttt{protocol/consensus/voter.go}
    \item \textbf{Modules:} \texttt{ChainedHotStuff.VoteRule()}, \texttt{NewChainedHotStuff()}
\end{itemize}

\paragraph{Root Cause Analysis}
This vulnerability represents the most extreme manifestation of "HighQC Ignorance." It exploits the discrepancy between the \texttt{HighQC} (the highest known certificate) and \texttt{bLock} (the locked block) during edge cases such as node recovery.
\begin{itemize}
    \item \textbf{Lock Amnesia:} The current implementation stores \texttt{bLock} in memory without persistence. Upon a node restart or crash-recovery, \texttt{bLock} is reset to \texttt{Genesis} (View 0).
    \item \textbf{Trivial Safety Satisfaction:} The Safety condition in \texttt{VoteRule} checks if a proposal extends the \texttt{bLock}. When \texttt{bLock = Genesis}, the \texttt{blockchain.Extends(block, Genesis)} call returns \texttt{true} for \textit{any} valid block in the system, effectively neutralizing the safety guardrail.
    \item \textbf{Liveness Bypass:} While the Liveness condition (\texttt{qcBlock.View > bLock.View}) typically prevents old forks, a leader can satisfy it using any QC with $\text{View} > 0$ (e.g., $QC_{B1}$) while simultaneously targeting nodes that have just reset their locks.
\end{itemize}

\paragraph{Attack Scenario: Full History Erasure}
Assume an established chain: $G \rightarrow B_1 \rightarrow B_2 \rightarrow B_3 \rightarrow B_4 \rightarrow B_5$ with $\text{HighQC} = QC_{B5}$.
\begin{enumerate}
    \item \textbf{Target Condition:} A subset of honest nodes restart, causing their \texttt{bLock} to revert to \texttt{Genesis}.
    \item \textbf{The Deep Fork:} A Byzantine leader ignores the entire history and proposes $B_6'$ extending directly from \texttt{Genesis} (Parent: \texttt{Genesis}, QC: $QC_{G}$).
    \item \textbf{Validation Failure:}
        \begin{itemize}
            \item \textbf{Liveness:} If using $QC_{B1}$, $QC_{B1}.View(1) > bLock.View(0)$, condition passes.
            \item \textbf{Safety:} \texttt{Extends($B_6'$, Genesis)} is \texttt{true}, condition passes.
        \end{itemize}
    \item \textbf{Outcome:} Honest nodes vote for $B_6'$, enabling the Byzantine leader to build an alternative chain ($G \rightarrow B_6' \rightarrow B_7' \dots$) that competes with the legitimate history, potentially leading to a massive state roll-back if the fork gains quorum.
\end{enumerate}

\paragraph{Impact \& Security Analysis}
This bug exposes the system to \textbf{Total History Erasure}. Unlike standard forks that deviate by a few blocks, this "Deep Fork" allows an attacker to rewrite the ledger from the Genesis block. While HotStuff's voting rules (one vote per view) prevent simultaneous double-commits within the same view, the "Lock Amnesia" creates a window where a Byzantine leader can force honest nodes to abandon the canonical chain in favor of a malicious alternative, causing severe consistency issues and service disruption.

\paragraph{Verification}
Reproduced via: \\
\texttt{go test -v -run TestByzantineLeaderBuildsDeepForkFromGenesis} \\
\textit{Result:}  \texttt{BUG CONFIRMED: Voter accepted proposal extending Genesis despite HighQC being at View 5.}
}

This bug is uncovered through further exploration of the bug described below.

\BugAnalysisBox{Bug: Design Vulnerability in Consensus Voting Logic}
{
\textbf{Case Study: Byzantine Leader Proposes Block Ignoring HighQC}

\paragraph{Affected Components}
\begin{itemize}
    \item \textbf{Files:} \texttt{protocol/rules/chainedhotstuff.go}, \texttt{protocol/consensus/voter.go}
    \item \textbf{Modules:} \texttt{ChainedHotStuff.VoteRule()}, \texttt{NewChainedHotStuff()}
\end{itemize}

\paragraph{Root Cause Analysis}
This issue stems from a design-implementation gap where \texttt{VoteRule} validates proposals against \texttt{bLock} (Locked Block) rather than the \texttt{HighQC} (Highest Quorum Certificate). Under specific edge cases—such as node recovery or state synchronization—\texttt{bLock} can become desynchronized from the actual chain height.

\begin{itemize}
    \item \textbf{Implementation Logic:} Per the HotStuff paper, a vote is cast if:
    \begin{enumerate}
        \item \textbf{Liveness:} \texttt{QC.View > bLock.View} 
        \item \textbf{Safety:} The block extends \texttt{bLock}.
    \end{enumerate}
    \item \textbf{Persistence Failure:} \texttt{bLock} is initialized to \texttt{Genesis} and kept only in memory. It is updated during the \texttt{CommitRule} but is not persisted to disk.
    \item \textbf{The Vulnerability:} Upon restart, a node's \texttt{bLock} reverts to \texttt{Genesis} (View 0). Even if the blockchain is recovered, the node will accept any proposal that satisfies the Liveness condition against a view of 0, even if the proposal ignores the actual canonical tip (\texttt{HighQC}).
\end{itemize}

\paragraph{Attack Scenarios}
\textbf{Scenario: Node Recovery / New Joiner}
\begin{enumerate}
    \item \textbf{State:} Chain exists as $G \rightarrow B_1 \rightarrow B_2 \rightarrow B_3$ (\texttt{HighQC} = $B_3$).
    \item \textbf{Trigger:} A node restarts; \texttt{bLock} resets to \texttt{Genesis}.
    \item \textbf{The Attack:} A Byzantine leader proposes $B_4'$ extending $B_1$ using $QC_{B1}$.
    \item \textbf{Validation:} \texttt{VoteRule} sees $QC_{B1}.View(1) > bLock.View(0)$. The Liveness condition passes, and the honest node mistakenly votes for a fork that ignores $B_3$.
\end{enumerate}

\paragraph{Verification}
Reproduced via: \texttt{go test -v -run TestByzantineLeaderIgnoresHighQC ./protocol/consensus/}
\begin{quote}
\texttt{ BUG DETECTED: Voter.Verify() accepted malicious proposal extending B1 instead of HighQC B3} \\
\texttt{ BUG CONFIRMED: Honest node voted for malicious proposal!} \\
\texttt{Result: This allows a Byzantine leader to fork the chain and split the quorum.}
\end{quote}

\paragraph{Impact \& Security Analysis}
While this vulnerability strictly follows the HotStuff paper's safety rules (preventing double-commits within the same view via \texttt{lastVotedView} checks), it creates significant \textbf{Liveness and Stability risks}:
\begin{itemize}
    \item \textbf{Chain Forking:} Malicious leaders can force honest nodes to vote on stale branches.
    \item \textbf{Quorum Fragmentation:} Voting power is split between the canonical chain and the fork, stalling finality.
    \item \textbf{Increased Latency:} Recovery from such "HighQC ignorance" requires extra views to re-synchronize the locked state.
\end{itemize}
}

\section{Bug Example for Five Classes}
\label{appendsec:bugclassexample}
\BugAnalysisBox{Recovery\&Execution Divergence}
{
\textbf{CommittedCommandVanishing (Loss of Committed State)}

\paragraph{Affected Components}
\begin{itemize}
    \item \textbf{File:} \texttt{src/epaxos/epaxos.go}
    \item \textbf{Line Numbers:} 1542--1556
    \item \textbf{Logic:} The \texttt{else} branch in the recovery handler where the leader receives no information from the quorum.
\end{itemize}

\paragraph{Root Cause Analysis}
The vulnerability arises because the recovery leader assumes an instance is "empty" or "unused" if it receives no records from its recovery quorum. When this happens, it erroneously proposes a \texttt{NO-OP} to finalize the instance. This logic fails to account for \textbf{Amnesia} (node restarts losing volatile state) or \textbf{Network Partitions} that isolate the only nodes holding the original \texttt{Commit} record.

\paragraph{Attack Scenario (5 Nodes: $R_0, R_1, R_2, R_3, R_4$)}
\begin{enumerate}
    \item \textbf{Fast Path Execution:} $R_0$ proposes $Cmd_A$ (\texttt{Key="k", Val=999}). Nodes $R_0, R_1, R_2$ accept it (Fast Quorum).
    \item \textbf{Partial Commit:} $R_0$ sends a \texttt{Commit} message \textbf{only} to $R_2$. $R_2$ marks the command as \texttt{Committed}.
    \item \textbf{Compound Failure:} 
        \begin{itemize}
            \item $R_0$ crashes permanently.
            \item $R_1$ crashes and reboots (Amnesia), wiping its log.
            \item $R_2$ is partitioned from the rest of the cluster.
        \end{itemize}
    \item \textbf{Flawed Recovery:} $R_3$ initiates recovery for the same instance and queries $\{R_1, R_3, R_4\}$. 
    \item \textbf{Voiding the Instance:} $R_1, R_3,$ and $R_4$ all report no record. $R_3$ incorrectly concludes the instance was never used and proposes \texttt{NO-OP}.
    \item \textbf{State Divergence:} $R_1, R_3, R_4$ commit and execute \texttt{NO-OP}. When the partition heals, $R_2$ has already executed $Cmd_A$, causing a permanent state divergence.
\end{enumerate}

\paragraph{Impact \& Risk Assessment}
\begin{itemize}
    \item \textbf{Outcome:} Violation of the "Stability" property and Linearizability.
    \item \textbf{Feasibility:} High in environments without strict persistence (Sync-to-Disk) and frequent network partitions.
    \item \textbf{Risk Level:} \textbf{Medium} (Requires specific failure interleaving).
\end{itemize}

\paragraph{Corrective Action}
The protocol should either:
\begin{itemize}
    \item Strictly prohibit Amnesia by requiring persistent WAL (Write-Ahead Logs).
    \item Enhance the recovery logic to require an explicit "Proof of Non-Existence" before proposing a \texttt{NO-OP}.
\end{itemize}
}

\BugAnalysisBox{Persistence\& Monotonicity Violations}
{
\textbf{Case Study: ExecutedStateAmnesia (Double Execution Vulnerability)}

\paragraph{Problem Description}
In the \texttt{efficient-epaxos} implementation, the \texttt{EXECUTED} status of an instance and the execution progress tracker (\texttt{ExecedUpTo}) are not persisted to stable storage. Upon a replica crash and recovery, the system loses its execution cursor, leading to the re-application of already committed commands to the state machine.

\paragraph{Affected Components}
\begin{itemize}
    \item \textbf{Files:} \texttt{src/epaxos/epaxos.go}, \texttt{src/epaxos/epaxos-exec.go}
    \item \textbf{Variables:} \texttt{ExecedUpTo} (Execution cursor), \texttt{inst.Status}
\end{itemize}

\paragraph{Root Cause Analysis}
The vulnerability stems from the volatility of the execution state:
\begin{enumerate}
    \item \textbf{Volatile Progress Tracking:} The \texttt{ExecedUpTo} array is stored purely in memory. On reboot, it resets to $-1$, losing the boundary between executed and pending commands.
    \item \textbf{Status Volatility:} The transition to \texttt{EXECUTED} (\texttt{epaxos-exec.go:128}) is never recorded via \texttt{recordInstanceMetadata}.
    \item \textbf{Asynchronous Gap:} Since execution happens asynchronously after the \texttt{COMMITTED} phase, the metadata update logic (which usually triggers on commit) fails to capture the final "executed" lifecycle of the instance.
\end{enumerate}

\paragraph{Attack Scenario (Double Execution)}
\begin{enumerate}
    \item \textbf{Initial State:} $R_0$ commits and executes $(0,0): \text{INC}(x)$, $(0,1): \text{INC}(x)$. State $x=2$, \texttt{ExecedUpTo[0]} = 1.
    \item \textbf{Crash \& Amnesia:} $R_0$ crashes. \texttt{ExecedUpTo} is wiped. 
    \item \textbf{Recovery:} $R_0$ reboots, restores instances $(0,0)$ and $(0,1)$ as \texttt{COMMITTED} from disk, but its memory-based cursor is $-1$.
    \item \textbf{Re-execution:} The \texttt{executeCommands} loop re-scans from index 0 ($ExecedUpTo + 1$) and applies the commands again.
    \item \textbf{Divergence:} Final state $x=4$, while non-crashing replicas remain at $x=2$.
\end{enumerate}

\paragraph{Impact \& Security Analysis}
\begin{itemize}
    \item \textbf{Outcome:} Violation of \textbf{Exactly-Once Semantics}.
    \item \textbf{Risk Level:} \textbf{Critical} for financial ledgers or non-idempotent state transitions.
    \item \textbf{Integrity:} Causes irreversible state divergence that the consensus protocol cannot automatically repair.
\end{itemize}

\paragraph{Verification}
Reproduced via: \texttt{go test -v src/epaxos/execedupto\_amnesia\_test.go} \\
\textit{Result:} \texttt{FAIL: State machine mismatch. Expected x=3, got x=6.}
}


\BugAnalysisBox{Dependency\&Topology Flaws}
{
\textbf{DependencyDivergence (Execution Order Inversion)}

\paragraph{Problem Description}
The recovery mechanism may fail to preserve established dependency relationships. If the "bridge node"—the only node that witnessed the causal link between two commands—undergoes amnesia (state loss), the recovery process may treat ordered commands as concurrent. This leads to an execution order inversion through deterministic tie-breaking, causing state machine divergence.

\paragraph{Affected Components}
\begin{itemize}
    \item \textbf{File:} \texttt{src/epaxos/epaxos.go}
    \item \textbf{Line Numbers:} 1471--1488
    \item \textbf{Logic:} The \texttt{handlePrepareReply} logic where dependencies are merged or updated during the recovery phase based on quorum responses.
\end{itemize}

\paragraph{Root Cause Analysis}
EPaxos relies on nodes recording dependencies ($deps$) to serialize conflicting commands. The bug occurs when the only evidence of a dependency (e.g., $Cmd_B$ depends on $Cmd_A$) exists on a node that subsequently fails and loses its memory-based log. During recovery, the remaining nodes—each only aware of one of the commands—reconstruct the dependency graph without the causal link, mistakenly concluding the commands are concurrent.

\paragraph{Attack Scenario (5 Nodes: $R_0, R_1, R_2, R_3, R_4$)}
\begin{enumerate}
    \item \textbf{Split Proposal:} $R_1$ proposes $Cmd_A$; $R_0$ proposes $Cmd_B$ (both target Key $x$).
    \item \textbf{Bridge Node Setup:} 
        \begin{itemize}
            \item $R_2$ processes $Cmd_A$ first, then $Cmd_B$. It records $Cmd_B.deps = \{Cmd_A\}$.
            \item $R_3$ only sees $PreAccept(Cmd_A)$; $R_4$ only sees $PreAccept(Cmd_B)$. Both record empty dependencies.
        \end{itemize}
    \item \textbf{Initial Execution:} $R_2$ receives \texttt{Commit} for both and executes $A \rightarrow B$. Final state at $R_2$: $x=2$.
    \item \textbf{The Wipe (Amnesia):} $R_0, R_1$ crash permanently. $R_2$ crashes and reboots with an empty log (Amnesia).
    \item \textbf{Flawed Recovery:} 
        \begin{itemize}
            \item $R_3$ recovers $Cmd_A$. Quorum $\{R_3, R_4, R_2\}$ reports no dependency for $A$ on $B$.
            \item $R_4$ recovers $Cmd_B$. Quorum $\{R_4, R_3, R_2\}$ reports no dependency for $B$ on $A$.
            \item The link $B \rightarrow A$ is lost. Both are committed with $deps = \{\}$.
        \end{itemize}
    \item \textbf{Execution Inversion:} The surviving nodes apply tie-breaking ($R_0.id < R_1.id$). $Cmd_B$ is executed before $Cmd_A$. Final state: $x=1$.
\end{enumerate}

\paragraph{Impact \& Risk Assessment}
\begin{itemize}
    \item \textbf{Outcome:} \textbf{Linearizability Violation}. The system history is rewritten, reversing the order of operations already visible to the user.
    \item \textbf{Risk Level:} \textbf{High} in high-concurrency environments with non-persistent deployments.
    \item \textbf{Feasibility:} Highly likely if network latency is asymmetric and nodes do not use synchronous disk writes for every \texttt{PreAccept}.
\end{itemize}

\paragraph{Corrective Action}
\begin{itemize}
    \item \textbf{Persistence:} Enforce stable storage for all metadata changes to prevent state loss upon reboot.
    \item \textbf{Recovery Logic:} Modify the recovery leader's criteria to require a "Strong Quorum" or "Epoch-based" validation to ensure that any previously committed dependency link is not ignored simply because a node is empty.
\end{itemize}
}

\BugAnalysisBox{Message Binding \& Signature Violations}
{
\textbf{Case Study: HighQC View Mismatch (View Tampering via UpdateHighQC)}

\paragraph{Problem Description}
The \texttt{UpdateHighQC} function checks if a block exists and if its view is higher than the current \texttt{HighQC}. However, it \textbf{fails to verify that the View stored within the QC matches the View of the block it certifies}. This allows an attacker to inject a QC with a tampered (artificially inflated) View field into the node's state.

\paragraph{Affected Components}
\begin{itemize}
    \item \textbf{File:} \texttt{protocol/viewstates.go}
    \item \textbf{Function:} \texttt{UpdateHighQC}
    \item \textbf{Dependencies:} \texttt{hotstuff.QuorumCert}, \texttt{blockchain} module
\end{itemize}

\paragraph{Root Cause Analysis}
The implementation logic compares the block's inherent view (\texttt{newBlock.View()}) for the threshold check but then blindly persists the entire \texttt{qc} object without consistency validation. 
\begin{itemize}
    \item \textbf{Consistency Gap:} There is no assertion that \texttt{qc.View() == newBlock.View()}.
    \item \textbf{Implicit Trust:} The function assumes the \texttt{QuorumCert} provided is internally consistent, even though the \texttt{View} field in the QC is not cryptographically bound to the signature in this specific implementation.
\end{itemize}

\paragraph{Attack Scenario (View Inflation)}
\begin{enumerate}
    \item \textbf{Setup:} The network is currently at \texttt{View v}.
    \item \textbf{Acquisition:} A Byzantine node obtains a valid \texttt{QC} certifying \texttt{Block\_A} from \texttt{View v-k}.
    \item \textbf{Tampering:} The attacker constructs a \texttt{fakeQC} with the original valid signature for \texttt{Block\_A} but an inflated \texttt{View: v + 1000}.
    \item \textbf{Replay:} The attacker sends this \texttt{fakeQC} to an honest node.
    \item \textbf{State Corruption:} Because \texttt{Block\_A.View()} exists and is valid, the node updates its \texttt{highQC} to \texttt{fakeQC}. The node now erroneously believes the highest certified view is \texttt{v + 1000}, causing it to reject legitimate messages from the actual current view.
\end{enumerate}

\paragraph{Impact \& Security Analysis}
\begin{itemize}
    \item \textbf{Liveness Violation:} Disrupts View Synchronization, locking honest nodes out of consensus.
    \item \textbf{Category:} \textbf{CWE-20: Improper Input Validation} (Cross-field consistency check missing).
    \item \textbf{Severity:} \textbf{High}. State-pollution that can be triggered remotely.
\end{itemize}

\paragraph{Corrective Action}
Add an explicit consistency check between the QC and the block metadata:
\begin{quote}
\texttt{if qc.View() != newBlock.View() \{ \\
\quad return false, fmt.Errorf("QC view mismatch") \\
\}}
\end{quote}
}


\BugAnalysisBox{Resource \& Operational Visibility Violations}
{
\textbf{Case Study: BlockManager Memory Exhaustion (OOM via Unbounded MissingBlocks)}

\paragraph{Problem Description}
The \texttt{BlockManager} maintains a \texttt{missing\_blocks} set to track parent blocks that have not yet been received. This data structure lacks an upper bound on its capacity, allowing a Byzantine validator to flood the system with blocks referencing non-existent ancestors. This leads to an Out-of-Memory (OOM) condition and potential Denial of Service (DoS).

\paragraph{Affected Components}
\begin{itemize}
    \item \textbf{Module:} \texttt{BlockManager}
    \item \textbf{Data Structure:} \texttt{missing\_blocks} (Set/Map)
    \item \textbf{Source Note:} Explicitly acknowledged by a \texttt{TODO} comment regarding Byzantine causal history and OOM risks.
\end{itemize}

\paragraph{Vulnerability Details}
\begin{enumerate}
    \item \textbf{Unbounded Growth:} There is no hard limit on the number of entries in the \texttt{missing\_blocks} collection. Attackers can continuously produce blocks with fake causal histories.
    \item \textbf{Future-Flooding Attack:} A naive "lowest round first" eviction strategy can be bypassed. Attackers can use high-round garbage blocks to force the eviction of legitimate, low-round missing blocks that the node actually needs.
\end{enumerate}

\paragraph{Attack Scenario (Future-Flooding)}
\begin{enumerate}
    \item \textbf{Normal State:} An honest node is at \texttt{Round 10} and is missing a legitimate parent block $P$ (from \texttt{Round 10}).
    \item \textbf{Flooding:} A Byzantine attacker floods the network with blocks referencing thousands of fake parent blocks at \texttt{Round 1000}.
    \item \textbf{Malicious Eviction:} If a capacity limit exists but uses a simple round-based eviction, the node will discard the legitimate \texttt{Round 10} entry to make room for the "newer" \texttt{Round 1000} entries.
    \item \textbf{Liveness Failure:} The honest node can never complete the causal history for \texttt{Round 10}, stalling its consensus progress.
\end{enumerate}

\paragraph{Impact \& Security Analysis}
\begin{itemize}
    \item \textbf{Vulnerability Type:} \textbf{CWE-770: Allocation of Resources Without Limits}.
    \item \textbf{Attack Cost:} Extremely low. A single committee member can trigger this with low-bandwidth flooding over time.
    \item \textbf{Consequence:} System-wide crash (OOM) or permanent stalling of honest nodes.
\end{itemize}

\paragraph{Corrective Action}
\begin{itemize}
    \item \textbf{Bound the Collection:} Implement a strict maximum size for \texttt{missing\_blocks}.
    \item \textbf{Smart Eviction:} Use a \textbf{Round-Proximity} strategy. Prioritize missing blocks closer to the current round and implement per-validator quotas.
\end{itemize}
}

\subsection{Final Summary}

\section{Prompt Example}
\label{appendsec:prompt}

\subsection{\agenta Prompt}

\begin{agentbox}[Orchestrator Agent]{
    \textbf{System:}
    \begin{Verbatim}[fontsize=\scriptsize]
You are the Orchestrator Agent, the central coordinator of a multi-agent system for discovering

vulnerabilities in distributed consensus protocols.

Your responsibilities:
1. Analyze confirmed bugs discovered by the Strategy Agent and TestGen Agent
2. Generate comprehensive bug reports with root cause analysis
3. Extract reusable vulnerability patterns for testing other protocol implementations
4. Track resource usage (tokens, cost) for each bug discovery

Be technical, precise, and actionable. Focus on providing insights that help developers understand

and fix the vulnerabilities.
    \end{Verbatim}
    
    \textbf{User:}
    
    \begin{Verbatim}[fontsize=\scriptsize]
As the Orchestrator Agent, analyze the following confirmed vulnerability and generate a comprehensive report.

## Confirmed Bug Details

The Strategy Agent generated an attack scenario, and the TestGen Agent confirmed it as a real vulnerability.

**Protocol**: {protocol_name}
**Protocol Type**: {protocol_type.upper()}
**Bug Name**: {scenario_name}
**Target Component**: {target_component or 'N/A'}
**Attack Category**: {attack_category or 'N/A'}

**Vulnerability Hypothesis**:
{vulnerability_hypothesis or 'N/A'}

**Preconditions**:
{preconditions_text}

**Attack Steps**:
{steps_text}

**Expected Bug Behavior**: {expected_bug_behavior or 'N/A'}
**Correct Behavior**: {correct_behavior or 'N/A'}

**Test File**: {test_file}
**Test Output** (excerpt):
```
{test_output[:2000] if test_output else 'N/A'}
```

## Resource Usage for This Bug Discovery

- **Prompt Tokens**: {prompt_tokens:,}
- **Completion Tokens**: {completion_tokens:,}
- **Total Tokens**: {total_tokens:,}
- **API Cost**: ${api_cost:.4f}

---

## Your Task

As the Orchestrator Agent, generate:

1. **Bug Report**: A comprehensive analysis including:
   - Root cause of the vulnerability
   - Potential security impact
   - Recommended fixes

2. **Bug Pattern**: A reusable pattern for testing similar vulnerabilities in other consensus protocol implementations.

## Output Format

<bug_report>
[Your detailed bug report with root cause analysis, impact assessment, and \\
fix recommendations]
</bug_report>

<bug_pattern>
Pattern Name: [A concise, descriptive name for this vulnerability pattern]
Bug Category: [safety/liveness/agreement]
Fault Type: [e.g., byzantine_leader, network_partition, crash_recovery]
Trigger Condition: [Specific conditions that trigger this vulnerability]
Description: [Detailed description of the vulnerability pattern and why it  occurs]
Test Template: [How to test for this pattern in other protocol implementations]
</bug_pattern>
    \end{Verbatim}
    
    \bigskip
    }
    \end{agentbox}

\subsection{\agentb Prompt}
\begin{agentbox}[Strategy Agent]{
    \textbf{System:}
    \begin{Verbatim}[fontsize=\scriptsize]
You are a distributed systems security researcher, specializing in discovering vulnerabilities in consensus

protocol implementations.

## Your Goal

For the target protocol, devise **cunning, creative, ORIGINAL** test scenarios to discover:
- **Safety Violations**: Different nodes commit different values for the same slot
- **Liveness Violations**: System cannot make progress
- **Agreement Violations**: Nodes disagree on the committed state

## Available Tool

You have access to ONE tool:

### get_pattern_details(pattern_id: str)
Query detailed information about a known bug pattern. Use this when you see a pattern_id 
in the bug pattern list that seems relevant and you want to learn more before deciding.

**Usage Guidelines**:
- Only call this for patterns you're genuinely considering using
- The pattern_id is a summary - if it already seems irrelevant, don't query it
- If the tool warns that a pattern was already used, you MUST create a substantially different attack vector

## Knowledge Base

Use your knowledge of:
- Distributed systems theory
- Consensus protocol design (Paxos, Raft, PBFT, HotStuff, Tendermint, etc.)
- Common implementation pitfalls
- Byzantine fault tolerance
- Network partitions and timing attacks
- State machine replication
- Crash recovery scenarios

## Repository Context (If Provided)

- Use it to ground your scenario in realistic components and constraints.
- Prefer components or behaviors that are likely to exist in the target repo.

## Bug Pattern Context (If Provided)

- You may receive a list of pattern IDs (which summarize the bug).
- Use `get_pattern_details(pattern_id)` to learn more about any interesting pattern.
- Select **at most one** relevant pattern and adapt it to the target protocol.
- **DO NOT reuse a pattern that has already been used** - the tool will warn you if it was.
- If no pattern fits or all have been used, create an **original** scenario.

## Thinking Approach

1. **Understand the Protocol Type**
   - What is the fault model? (CFT vs BFT)
   - What are the core safety/liveness assumptions?
   - What are the key invariants that must hold?

2. **Identify Attack Vectors**
   Think about common vulnerability categories:
   - Race conditions in concurrent message handling
   - State persistence and recovery bugs
   - View change / leader election edge cases
   - Quorum intersection violations
   - Timeout and timing-related issues
   - Message ordering and delivery assumptions
   - Equivocation and double-voting scenarios
   - Lock state management errors

3. **Creative Scenario Design**
   - Consider edge cases that developers might miss
   - Think about interactions between multiple components
   - Imagine adversarial network conditions
   - Consider crash-restart scenarios at critical moments

4. **Design Verification**
   - How do we determine if the attack succeeded?
   - What specific assertions can capture this issue?

## IMPORTANT: Avoid Unrealistic Scenarios

Before proposing any attack scenario, verify it is REALISTIC:

1. Can the trigger condition actually happen in a real deployment?
2. Are there upstream checks that prevent the scenario?
3. Is the fault model (CFT vs BFT) correctly applied?

Examples of UNREALISTIC scenarios to avoid:
- Testing CFT protocol with Byzantine behavior (CFT assumes honest nodes)
- Assuming attacker can bypass all input validation
- Assuming network delivers messages in impossible orders
- Creating test states that cannot exist in normal operation

**A scenario is only valid if it can genuinely occur in production!**

## Output Format

<thinking>
[Understanding the Protocol]
...I observe that the key mechanism of this protocol is...
...It assumes...
...The fault model is...

[Identifying Attack Vectors]
...Common vulnerabilities in this type of protocol include...
...I should consider...

[Creative Reasoning]
...If I construct a scenario where...
...This might lead to...
...The key insight is...

[Designing Verification]
...If the attack succeeds, I should observe...
...The assertion should check...
</thinking>

<attack_scenario>
Name: [Descriptive name, e.g., "CrashInducedLockAmnesiaLeadingToConflictingCommits"]

Target Component: [Specific component/function being attacked]

Vulnerability Hypothesis: [What problem you believe exists]

Attack Category: [safety / liveness / agreement]

Inspiration Source: [pattern_memory::<pattern_id> | original]

Preconditions:
1. [What initial state is needed]
2. [...]

Attack Steps:
1. [Specific action]
2. [...]
3. [...]

Expected Bug Behavior: [What will happen if there's a bug]

Correct Behavior: [How the protocol should respond]

Assertions:
1. [How to verify the attack succeeded]
2. [...]
</attack_scenario>
    \end{Verbatim}
    
    \textbf{User:}
    
    \begin{Verbatim}[fontsize=\scriptsize]
## Task

Devise a **cunning, creative, ORIGINAL** attack scenario for {protocol_name}.

Use your knowledge of distributed systems and consensus protocols to identify potential vulnerabilities.
Think about:
- What could go wrong in the implementation?
- What edge cases might developers miss?
- How could an adversary (or faulty node) exploit timing or state management?

**Remember**: 
- Ensure your scenario is REALISTIC and can genuinely occur in production!
- Be CREATIVE and think of scenarios that might not be obvious
- Focus on implementation-level bugs, not protocol design flaws

Now think deeply about potential vulnerabilities and design an attack scenario...
    \end{Verbatim}
    \textbf{Bug Exploitation}
    
    \begin{Verbatim}[fontsize=\scriptsize]
You are a distributed systems security researcher in BUG EXPLOITATION MODE.

A bug has been CONFIRMED in the target protocol. Your mission is to leverage this 
discovery to uncover MORE vulnerabilities by exploring related attack vectors.

## Exploitation Strategies

When exploiting a confirmed bug, consider these approaches:

### 1. Root Cause Analysis
- What is the fundamental assumption being violated?
- Where else in the codebase might this assumption be violated?
- Are there other code paths that share the same flawed logic?

### 2. Variant Generation
- Can the same bug be triggered through different entry points?
- What happens with different timing? Different message orders?
- Are there edge cases the original test didn't cover?

### 3. Pattern Propagation
- If this component has this bug, do similar components have it too?
- Does the bug exist in related protocols (view change, leader election, etc.)?
- Can the bug be combined with other faults to cause more severe issues?

### 4. Deeper Exploration
- What happens if we push the bug further?
- Can we escalate from liveness to safety violations?
- Are there cascading effects in dependent components?

## Key Principles

1. **Build on the Discovery**: Use the confirmed bug as a foundation
2. **Think Creatively**: The bug reveals a blind spot - where else might it exist?
3. **Avoid Repetition**: Generate genuinely NEW scenarios, not minor variations
4. **Stay Realistic**: All scenarios must be achievable in real deployments

## Output Format

<thinking>
[Analyzing the Confirmed Bug]
...The root cause of this bug is...
...This reveals an assumption that...

[Identifying Related Attack Vectors]
...Based on this, I should explore...
...A similar vulnerability might exist in...

[Designing New Scenario]
...If I target X component with Y approach...
...This is different from the original because...
</thinking>

<attack_scenario>
Name: [New descriptive name - must be different from original]
Target Component: [May be same or different component]
Vulnerability Hypothesis: [Related but distinct vulnerability]
Attack Category: [safety / liveness / agreement]
Relation to Original Bug: [How this relates to the confirmed bug]

Preconditions:
1. [Required conditions]
2. [...]

Attack Steps:
1. [Specific action]
2. [...]

Expected Bug Behavior: [What happens if this variant exists]
Correct Behavior: [Expected correct protocol response]

Assertions:
1. [Verification method]
2. [...]
</attack_scenario>
    \end{Verbatim}
    
    \bigskip
    }
\end{agentbox}
\bigskip
\subsection{\agentc Prompt}
\begin{agentbox}[TestGen Agent]{
    \textbf{System:}
    \begin{Verbatim}[fontsize=\scriptsize]
You are a test code generation expert, skilled at transforming attack scenarios into executable test code.

## Your Goal

Transform the attack scenario generated by Strategy Agent into:
1. **Repository-style compliant** test code
2. **Executable, verifiable** test cases
3. **Clear assertions** to determine if the attack succeeded

## Available Tools for Repository Knowledge

You have these powerful tools to leverage accumulated knowledge:

### `repo_knowledge(compact=True)`
Get cached knowledge about the repository. Use `compact=True` (default) for LLM-optimized summary.
- Contains: test structure, coding style, helper functions with signatures and usage examples
- Contains: lessons learned from previous errors

### `repo_knowledge_find_lessons(error_message)`
**CALL THIS FIRST** when encountering compilation or test errors.
- Searches previous lessons that match the error
- Returns solutions that worked before

### `repo_knowledge_add_lesson(issue, solution, error_pattern="")`
**ALWAYS CALL THIS** after successfully fixing an error.
- Records the issue and solution for future reference
- `error_pattern`: Optional regex to match similar errors

### `repo_knowledge_add_helper(name, file, purpose, signature="", usage_example="", returns="")`
Call this when you discover a useful helper function.
- Records detailed information including signature and usage example
- Helps future test generation use the helper correctly

## Workflow

1. **Analyze Repository Structure**
   - Call `repo_knowledge()` to get test structure, coding style, and helper functions
   - Look for helpers with `signature` and `usage_example` - use them directly
   - Check `lessons_learned` for common pitfalls to avoid
   - Find similar test files as reference if needed

2. **Generate Test Code**
   - Follow the repository's coding style exactly
   - Use helper functions with their documented signatures
   - Use correct package/module declarations
   - Import necessary dependencies

3. **Add Assertions**
   - Detect expected vulnerability behavior
   - If protocol is correct, test should PASS
   - If there's a bug, test should FAIL

4. **Execute Test**
   - Write the test file
   - Run the test command
   - Collect output results

## Code Quality Requirements

- **Correct package name**: Read from existing files, don't guess
- **Correct imports**: Use modules that actually exist in the repository
- **Follow style**: Mimic how other tests in the repository are written
- **Clear comments**: Explain what each step is doing

## Error Handling Workflow (CRITICAL)

When test compilation/execution fails, follow this **exact** sequence:

### Step 1: Search Previous Lessons FIRST
```
error_message = "..." # the error you got
-> Call repo_knowledge_find_lessons(error_message)
```
If a relevant lesson is found, apply the solution directly.

### Step 2: Investigate If No Lesson Found
- Use Bash to run commands: `ls`, `grep`, etc.
- Read relevant files to understand the issue
- Check if imports, package names, paths are correct

### Step 3: Fix and Verify
- Make corrections based on your investigation
- Run the test again

### Step 4: RECORD THE LESSON (Don't forget!)
After successfully fixing:
```
-> Call repo_knowledge_add_lesson(
    issue="What went wrong (be specific)",
    solution="How you fixed it (be detailed)",
    error_pattern="optional regex to match similar errors"
)
```

### Step 5: Update Helper Details If Discovered
If you found a helper function's signature or usage during investigation:
{helper_example}

**Key Principle**: Learn from errors and RECORD what you learn. 
Every error you fix should benefit future test generation.

## Output Format

<analysis>
[Analyzing repository test structure]
...repo_knowledge shows test files are in...
...Naming convention is...
{analysis_example}
...Previous lessons: avoid X, do Y instead...
</analysis>

<test_code language="{lang_order}">
// Complete test code (see language-specific templates below)
</test_code>

<execution_plan>
File Path: [Path where test file will be written]
Test Command: [Command to run the test]
</execution_plan>
    \end{Verbatim}
    
    \textbf{User:}
    
    \begin{Verbatim}[fontsize=\scriptsize]
## Language-Specific Template

{language_template}
"""
    )

    prompt_parts.append(
        f"""
## Target Protocol

- **Name**: {protocol_name}
- **Language**: {language}

## Attack Scenario

{attack_scenario}
"""
    )

    if repo_knowledge:
        prompt_parts.append(
            f"""
## Repository Knowledge (Pre-loaded)

{repo_knowledge}
"""
        )

    # Add test commands
    test_commands = get_test_commands(language)
    prompt_parts.append(test_commands)

    prompt_parts.append(
        """
## Task

1. Review the repo_knowledge above (especially helper signatures and lessons learned)
2. Use the language-specific template as a guide for code structure
3. Transform the attack scenario into test code using available helpers
4. Ensure code can compile and run
5. Execute test and report results
6. If errors occur: search lessons -> fix -> record lesson

Now begin...
    \end{Verbatim}
    
    \bigskip
    \textbf{Go Test Template}
    
    \begin{Verbatim}[fontsize=\scriptsize]
## Go Test Template

```go
package <package_name>

import (
    "testing"
    "time"
    // Add other imports from repo_knowledge
)

// Test<ScenarioName> tests <brief description>
//
// Attack scenario: <scenario description>
// Expected behavior: <what the test verifies>
func Test<ScenarioName>(t *testing.T) {
    // Step 1: Setup cluster/nodes
    // Use helpers from repo_knowledge (e.g., NewNetwork, MakeCluster)
    
    // Step 2: Establish initial state
    // - Elect leader if needed
    // - Submit baseline transactions
    
    // Step 3: Execute attack scenario
    // - Create partition / inject faults
    // - Trigger specific protocol conditions
    
    // Step 4: Verify safety/liveness properties
    // - Check for safety violations
    // - Verify expected behavior
    
    // Step 5: Cleanup and final assertions
    if safetyViolation {
        t.Fatalf("SAFETY VIOLATION: %s", details)
    }
}
```

### Go Test Commands
- Single test: `go test -v -run TestName ./path/...`
- With timeout: `go test -v -timeout 60s -run TestName ./...`
- With race detector: `go test -v -race -run TestName ./...`
    \end{Verbatim}
    \textbf{Rust Test Template}
    
    \begin{Verbatim}[fontsize=\scriptsize]
## Rust Test Template

```rust
#[cfg(test)]
mod tests {
    use super::*;
    // Add imports from repo_knowledge
    
    /// Creates a logger for test output
    fn default_logger() -> Logger {
        let decorator = slog_term::TermDecorator::new().build();
        let drain = slog_term::FullFormat::new(decorator).build().fuse();
        let drain = slog_async::Async::new(drain).build().fuse();
        Logger::root(drain, o!())
    }
    
    /// Test<scenario_name> tests <brief description>
    ///
    /// Attack scenario: <scenario description>
    /// Expected behavior: <what the test verifies>
    #[test]
    fn test_<scenario_name>() {
        let logger = default_logger();
        
        // Step 1: Setup cluster/nodes
        // Use helpers from repo_knowledge (e.g., Network::new, RawNode::new)
        
        // Step 2: Establish initial state
        // - Trigger elections
        // - Propose baseline entries
        
        // Step 3: Execute attack scenario
        // - Simulate partitions
        // - Inject faults
        
        // Step 4: Verify safety/liveness properties
        assert!(condition, "SAFETY VIOLATION: {}", details);
        
        // Step 5: Cleanup and final assertions
        println!("Test passed - no safety violations detected");
    }
}
```

### Rust Test Commands
- Single test: `cargo test test_name -- --nocapture`
- In harness: `cargo test -p harness test_name -- --nocapture`
- With logging: `RUST_LOG=debug cargo test test_name -- --nocapture`
    \end{Verbatim}
    
    \textbf{Java Test Template}
    
    \begin{Verbatim}[fontsize=\scriptsize]
## Java Test Template (JUnit 5)

```java
package <package.name>;

import org.junit.jupiter.api.Test;
import org.junit.jupiter.api.BeforeEach;
import org.junit.jupiter.api.AfterEach;
import static org.junit.jupiter.api.Assertions.*;
// Add other imports from repo_knowledge

/**
 * Test<ScenarioName> - <brief description>
 *
 * Attack scenario: <scenario description>
 * Expected behavior: <what the test verifies>
 */
public class <ScenarioName>Test {

    // Test fixtures
    private TestCluster cluster;
    
    @BeforeEach
    public void setUp() throws Exception {
        // Initialize test cluster
        // Use helpers from repo_knowledge
    }
    
    @AfterEach
    public void tearDown() throws Exception {
        // Cleanup resources
        if (cluster != null) {
            cluster.shutdown();
        }
    }
    
    @Test
    public void test<ScenarioName>() throws Exception {
        // Step 1: Setup
        // Create cluster with required configuration
        
        // Step 2: Establish initial state
        // - Wait for leader election
        // - Submit baseline transactions
        
        // Step 3: Execute attack scenario
        // - Create partition / inject faults
        // - Submit conflicting transactions
        
        // Step 4: Verify safety properties
        assertFalse(safetyViolation, "SAFETY VIOLATION: " + details);
        
        // Step 5: Heal partition and verify convergence
        assertEquals(expectedState, actualState, "States should converge after partition heals");
    }
}
```

### Java Test Commands
- Maven: `mvn test -Dtest=TestClassName`
- Maven single method: `mvn test -Dtest=TestClassName#testMethodName`
- Gradle: `gradle test --tests TestClassName`
\end{Verbatim}
}
\end{agentbox}

\end{document}